\documentclass[a4paper,fleqn,usenatbib]{mnras}

\usepackage{newtxtext,newtxmath}

\usepackage[T1]{fontenc}
\usepackage{ae,aecompl}

\usepackage{graphicx}	
\usepackage{amsmath}	
\usepackage{amssymb}	
\usepackage{threeparttable}
\usepackage{multirow}

\newcommand{\eff}{\epsilon_{\mathrm{ff}}}
\newcommand{\effz}{\epsilon_{\mathrm{ff,0}}}
\newcommand{\effobs}{\epsilon_{\mathrm{ff,obs}}}
\newcommand{\tff}{t_{\mathrm{ff}}}
\newcommand{\tffz}{t_{\mathrm{ff,0}}}
\newcommand{\tsf}{t_{\mathrm{sf}}}
\newcommand{\tc}{t_{\mathrm{coll}}}
\newcommand{\tfb}{t_{\mathrm{fb}}}
\newcommand{\Mg}{{M_\mathrm{g}}}
\newcommand{\Mgz}{{M_\mathrm{g,0}}}
\newcommand{\dMg}{{\dot{M}_\mathrm{g}}}
\newcommand{\Ms}{{M_\mathrm{*}}}
\newcommand{\dMs}{{\dot{M}_\mathrm{*}}}

\newcommand{\dMa}{{\dot{M}_\mathrm{acc}}}

\newcommand{\dMf}{{\dot{M}_\mathrm{fb}}}
\newcommand{\krho}{{k_\rho}}

\newcommand{\ta}{t_{\mathrm{acc}}}
\newcommand{\taua}{\tau_{\mathrm{acc}}}
\newcommand{\taufb}{\tau_{\mathrm{fb}}}
\newcommand{\tobs}{t_{\mathrm{clust}}}
\newcommand{\tauobs}{\tau_{\mathrm{clust}}}

\newcommand{\tauc}{\tau_{\mathrm{coll}}}
\newcommand{\phid}{\phi_{\mathrm{d}}}
\newcommand{\SFR}{\mathrm{SFR}}

\newcommand{\beq}	{\begin{equation}}
\newcommand{\eeq}	{\end{equation}}
\newcommand{\beqa}{\begin{eqnarray}}
\newcommand{\eeqa}{\end{eqnarray}}
\newcommand\caln{{\cal N}}
\newcommand{\dis}{\displaystyle}

\newcommand{\pbyp}[1]	{{{\partial\hfil}\over{\partial#1}}}
\newcommand{\ppbyp}[2]	{{{\partial#1}\over{\partial#2}}}
\def\simlt{\lower.5ex\hbox{$\; \buildrel < \over \sim \;$}}
\def\simgt{\lower.5ex\hbox{$\; \buildrel > \over \sim \;$}}

\def\ga{\simgt}

\newcommand{\aref}[1]{\hyperref[#1]{Appendix~\ref{#1}}}

\usepackage{color}

\title[Bound star cluster formation]{How do bound star clusters form?}

\author[Krumholz \& McKee]{
Mark R. Krumholz$^{1,2}$\thanks{E-mail: mark.krumholz@anu.edu.au (MRK)}
and Christopher F. McKee$^{3}$
\\
$^{1}$Research School of Astronomy and Astrophysics, Australian National University, Canberra, ACT 2611 Australia\\
$^{2}$ARC Centre of Excellence for Astronomy in Three Dimensions (ASTRO-3D), Canberra, ACT 2611 Australia\\
$^{3}$Departments of Physics and Astronomy, University of California, Berkeley, CA 94720 USA
}

\date{Accepted XXX. Received YYY; in original form ZZZ}

\pubyear{2019}

\begin{document}
\label{firstpage}
\pagerange{\pageref{firstpage}--\pageref{lastpage}}
\maketitle

\begin{abstract}
Gravitationally-bound clusters that survive gas removal represent an unusual mode of star formation in the Milky Way and similar spiral galaxies. While forming, they can be distinguished observationally from unbound star formation by their high densities, virialised velocity structures, and star formation histories that accelerate toward the present, but extend multiple free-fall times into the past. In this paper we examine several proposed scenarios for how such structures might form and evolve, and carry out a Bayesian analysis to test these models against observed distributions of protostellar age, counts of young stellar objects relative to gas, and the overall star formation rate of the Milky Way. We show that models in which the acceleration of star formation is due either to a large-scale collapse or a time-dependent increase in star formation efficiency are unable to satisfy the combined set of observational constraints. In contrast, models in which clusters form in a ``conveyor belt'' mode where gas accretion and star formation occur simultaneously, but the star formation rate per free-fall time is low, can match the observations. 
\end{abstract}

\begin{keywords}
galaxies: star clusters: general -- ISM: kinematics and dynamics -- open clusters and associations: general -- stars: formation
\end{keywords}




\section{Introduction}
\label{sec:intro}

The typical outcome of star formation in spiral galaxies is not a gravitationally-bound star cluster. In the Milky Way, \citet{lada03a} were among the first to point out that the number of observed star clusters at ages from $10-100$ Myr is a factor of $\sim 10$ smaller than one would expect if every observed gas-embedded star-forming clump were to go on to become a cluster of comparable mass. The natural explanation for this discrepancy is that most of the young stars that we observe in star-forming regions are in fact unbound, or will become so once the gas is removed, and that we count them as cluster members at young ages simply because they have not yet had time to drift apart. Extensive surveys of external galaxies echo this conclusion, with counts of star clusters as a function of age implying that no more than $5-10\%$ of stars that form will remain part of a gravitationally-bound structure several tens of Myr after formation (e.g., \citealt{adamo15a, johnson16a, chandar17a, messa18a}; see the recent review by \citealt{krumholz19a} for additional references).

Thus regions of star-formation that do go on to become gravitationally-bound clusters must be special in some way. Recent observational advances offer significant hints about how such regions might be special. Regions that go on to become bound clusters do not appear to represent a distinct class of gas cloud, such that most clouds unbind entirely and a small minority remain mostly bound. Instead, many star-forming regions appear to consist of a dense inner part that contains a minority of the mass, which is likely to go on to become bound, and an extended outer part whose stars will drift apart. The inner regions that go on to become bound are distinguishable in several ways.

First, they appear to feature extended star-formation histories. Low-density star-forming regions that are $\sim 10$ pc in size or larger tend to have stellar populations whose ages are comparable to their crossing times \citep{elmegreen00a, kruijssen19a}, suggesting a relatively rapid formation process. By contrast, the densest regions of star formation, with sizes $\sim 1$ pc, have star formation histories that are significantly more extended compared to their dynamical times. The best-studied example is the Orion Nebula Cluster (ONC), where the free-fall time in the central 1 pc is $\approx 0.6$ Myr \citep{da-rio14a}, but there is extensive evidence that star formation has been ongoing for a significantly longer period \citep[e.g.,][]{reggiani11a, jaehnig15a, da-rio16a, beccari17a}. Star formation in this region appears to be accelerating \citep{palla00a, huff06a, caldwell18a}, but even accounting for this effect most stars are significantly older than a free-fall time -- using the kinematically-selected sample and estimated ages of \citet{kounkel18a}, \citet{krumholz19a} find that 50\% of the stars in the ONC are older than 3 free-fall times, and 10\% are older than 10 free-fall times. However, the ONC appears to be typical in this regard: similarly extended but accelerating star formation histories have been observed in NGC 6530 \citep[more than 25\% of stars older than 3 free-fall times]{prisinzano19a}, Perseus \citep{azimlu15a}, Taurus, and $\rho$ Ophiuchus \citep{caldwell18a}, though the last three of these regions are still highly gas-dominated, and it is therefore unclear if they will in fact reach star formation efficiencies sufficient to produce a bound cluster.

Second, the regions with extended star formation histories are also distinct kinematically. While most young stars still embedded in their parent molecular clouds are characterised by unrelaxed density and velocity distributions \citep[e.g.,][]{furesz08a, tobin09a}, the density distribution in the central 1 pc of the ONC can be fit reasonably well by an isothermal, spherically-symmetric \citet{king62a} model \citep{hillenbrand98a}, and the velocity distribution is virialised \citep{kim19a}. This region is neither expanding or contracting, and there is no evidence for a population of stars on primarily-radial orbits that are plausibly falling toward or escaping from it \citep{ward18a, kuhn19a}.

While regions like the ONC appear to be distinct in some respects, they also share one very significant commonality with the more extended envelopes around them. The density of young stellar objects (YSOs) increases smoothly with gas surface density, with no clear breaks at the densities or radii that correspond to the shift from unrelaxed, fractal stellar distributions to relaxed, virialised ones \citep{gutermuth11a}. Once one normalises the gas surface density by the free-fall time, it correlates remarkably tightly with YSO count; there is a near-linear relationship between YSO mass and gas mass normalised by free-fall time with a scatter of only $\approx 0.3-0.4$ dex across orders of magnitude in mass and density (\citealt{krumholz12a, lada13a, evans14a, heyer16a, ochsendorf17a} -- see Figure 10 of \citealt{krumholz19a} for a compilation of results). One can interpret this correlation as describing the efficiency of star formation: the star formation efficiency per free-fall time is $\epsilon_{\rm ff} = \dot{M}_* / (\Mg/\tff)$, where $\Mg$ and $\tff$ are the gas mass and free-fall time. If there are $N_{\rm YSO}$ YSOs associated with this gas that have a mean mass $M_{\rm YSO}$ and that remain spectrally-identifiable as such for a time $t_{\rm YSO}$, then the star formation rate must be $\dot{M}_* \approx N_{\rm YSO} M_{\rm YSO} / t_{\rm YSO}$.
All published studies based on YSO counts give $\eff \approx 0.01$, with $\lesssim 0.4$ dex scatter; the low value of $\eff$ and the extended star formation histories in regions that become bound are likely related, since a low $\eff$ region is likely to become bound only if it forms stars long enough to reach a respectable total star formation efficiency, and for the stars formed to dynamically relax \citep{kruijssen12a}. In contrast, ratios of far-infrared or free-free luminosity to gas mass give a much larger dispersion \citep{vutisalchavakul16a, lee16a, ochsendorf17a}. However, these results depend critically upon the procedure used to match regions of FIR or free-free emission to spatially-separated molecular clouds, with differing matching procedures yielding results that differ by up to $\sim 1$ dex \citep{krumholz19a}. Given the consistency of the much more direct YSO results, we regard them as more reliable.

Since regions like the ONC appear to be distinct from other star-forming regions in some ways but not others, and appear to evolve distinctly from the bulk of the young stellar population once star formation ends and gas is cleared, it is interesting to attempt to characterise the star formation process in these regions. Our goal in this paper is to examine a variety of proposed scenarios for star cluster formation that may be found in the literature, construct simple mathematical descriptions for them, and confront them with the wide variety of observational results that we have just outlined. We present the models to which we are interested in comparing, and outline a general framework for describing them, in \autoref{sec:models}. In \autoref{sec:obs} we compare these models to the observations outlined above, determining where they succeed and where they fail. We summarise our findings in \autoref{sec:conclusion}.


\section{Framework for cluster formation}
\label{sec:models}

We now sketch out some simple, general models for how star clusters might form. Before beginning this exercise, it is important to understand that our goal is not to examine fully self-consistent and detailed models for star cluster formation. Even purely analytic or semi-analytic models for cluster formation and cloud evolution \citep[e.g.,][]{goldbaum11a, zamora-aviles12a, zamora-aviles14a, lee16a, lee16c} generally include complex prescriptions for the time evolution of cloud mass, density, velocity dispersion, star formation activity, the effects of stellar feedback, and similar details. Comparing observations to such models is in general very difficult, because the models have many moving parts and contain numerous tuneable parameters. Our goal instead is to develop cartoons that capture some of the main qualitative features of models that have been proposed in the literature, but that are analytically-computable and have relatively few free parameters, so that we can carry out statistical comparisons to observation. This means that we will simply prescribe the evolution of parameters such as cloud mass and density, rather than trying to compute them fully self-consistently. As we introduce the individual models below, we will point out features of the more complex published models they are intended to capture.

All the software used to produce all the plots and analysis found in this paper are publicly available at \url{https://bitbucket.org/krumholz/km19/}.

\subsection{General framework}

We begin by characterising a gas cloud that is in the process of forming a star cluster in terms of its instantaneous gas mass $\Mg$ and mean density $\rho$; it is convenient to characterise the latter in terms of the free-fall time $\tff = \sqrt{3\pi/32 G \rho}$. Both $\Mg$ and $\rho$ can in general be functions of time. At any instant, the cloud forms stars at a rate 
\begin{equation}
\label{eq:sfr}
\dMs = \eff \frac{\Mg}{\tff}.
\end{equation}
For simplicity we will generally only worry about mean quantities, but we note that, if instead of a uniform cloud one considers a cloud where the density profile is a powerlaw $\rho\propto r^{-\krho}$, and one assumes that \autoref{eq:sfr} holds locally (i.e., at every point the star formation density obeys $\dot{\rho}_* = \eff \rho/\tff$), then the sole modification to \autoref{eq:sfr} is that $\eff$ is increased by a factor of $[2/(2-\krho)][(3-\krho)/3]^{3/2}$, which is of order unity unless $\krho$ is very close to 2.

In addition to star formation, the cloud can gain mass by accretion and lose it by ejection of mass by stellar feedback. We take the mass removal rate by feedback to be proportional to the star formation rate $\dMf = \eta \dMs$, while the accretion rate $\dMa$ is an input parameter; here $\eta$ is the usual mass loading factor.\footnote{Our choice to parameterise mass loss in terms of a mass-loading factor $\eta$, so that the mass removal rate is proportional to the star formation rate, differs from some other simple models \citep[e.g.,][]{lee16a} in which the mass removal rate is taken to be proportional to the total stellar mass. As discussed in \citet{dekel13a}, which of these approximations is preferable depends on how the duration of star formation compares to the duration of the feedback mechanisms that dominate mass removal -- $\dMf \propto \dMs$ is preferable if star formation is extended compared to feedback, $\dMf \propto M_*$ if not. The dominant feedback mechanisms in a forming star cluster are likely to be protostellar outflows (on for $\approx 0.1$ Myr) for clusters that do not contain O stars, and photoionisation or radiation pressure (on for $\approx 3$ Myr) for those that do \citep{krumholz19a}. Below we will compare to data on two star clusters, NGC 6530 and the ONC. In NGC 6530, the duration of star formation is $\approx 1-2$ Myr, and there are no O stars; in the ONC, there is an O star, but the duration of star formation is $\approx 3-4$ Myr. Since both of these systems have star formation durations comparable to or longer than the corresponding feedback duration, we prefer to model the mass removal rate as proportional to the instantaneous star formation rate.} The total mass of gas and stars therefore evolve following
\begin{equation}
\dMg = \dMa - \left(1+\eta\right) \eff \frac{\Mg}{\tff},
\qquad 
\dMs = \eff \frac{\Mg}{\tff}.
\label{eq:evol_eq}
\end{equation}
In principle both $\dMa$ and $\eta$ can, like $\tff$, be a function of time. 

\subsection{Scenarios of star formation}
\label{ssec:scenarios}

\begin{table*}
\caption{
\label{tab:model_list}
Summary of models and their parameters. Note that not all of these parameters are independent, and in cases where parameters are related, we list the relationship in the table.
}
\begin{tabular}{cccc}
\hline\hline
Model name & Abbreviation & Parameter & Meaning \\ \hline
\multicolumn{2}{c}{\multirow{4}{*}{Parameters common to all models}} & $\eff$ & Star formation efficiency per free-fall time \\
& & $\eta$ & Mass loading factor \\ 
& & $\tff$ & Free-fall time \\
& & $\tsf$ & Star formation timescale, $\tsf= \tff / [(1+\eta) \eff]$ \\
\hline
Static cloud & ST & -- & \\ \hline
\multirow{2}{*}{Conveyor belt} & \multirow{2}{*}{CB} & $p$ & Accretion rate versus time $\dMa \propto t^p$ \\
& & $\ta$ & Duration of accretion flow; dimensionless time $\taua \equiv \ta/\tsf$ \\ \hline
\multirow{3}{*}{Conveyor belt + dispersal} & \multirow{3}{*}{CBD} & $p$ & Accretion rate versus time $\dMa \propto t^p$ \\
& & $\ta$ & Duration of accretion flow; dimensionless time  $\taua \equiv \ta/\tsf$ \\ 
& & $\phid$ & Ratio of $1+\eta$ during dispersal phase to value during accretion phase \\
\hline
\multirow{3}{*}{Global collapse} & \multirow{3}{*}{GC} & $\tc$ & Collapse time; dimensionless time $\tauc \equiv \tc/\tsf$ \\
& & $\tffz$ & Free-fall time at onset of star formation; for this model $\tsf \equiv \tffz/[(1+\eta)\eff]$ \\
& & $\xi$ & Ratio of collapse timescale to free-fall timescale, $\tc = 2\tffz/\xi$ \\ 
\hline
\multirow{5}{*}{{Global collapse + dispersal}} & \multirow{5}{*}{{GCD}} 
& {$\tc$} & {Collapse time; dimensionless time $\tauc \equiv \tc/\tsf$} \\
& & {$\tffz$} & {Free-fall time at onset of star formation; for this model $\tsf \equiv \tffz/[(1+\eta)\eff]$} \\
& & {$\xi$} & {Ratio of collapse timescale to free-fall timescale, $\tc = 2\tffz/\xi$} \\ 
& & {$\tfb$} & {Time at which feedback increases; dimensionless $\taufb \equiv \tfb/\tsf$} \\
& & {$\phid$}  & {Ratio of $1+\eta$ during dispersal phase to value during earlier phase} \\
\hline
\multirow{3}{*}{Increasing efficiency} & \multirow{3}{*}{IE} & $\delta$ & Efficiency per free-fall time varies as $\eff = \effz (t/\tff)^\delta$ \\
& & $\effz$ & Value of $\eff$ at $t = \tff$; for this model, $\tsf \equiv \tff/[(1+\eta)\effz]$ \\
& & $\chi$ & Ratio of star formation timescale to free-fall timescale, $\chi = \tsf/\tff$
\\ \hline\hline
\end{tabular}
\end{table*}

\begin{figure}
\includegraphics[width=\columnwidth]{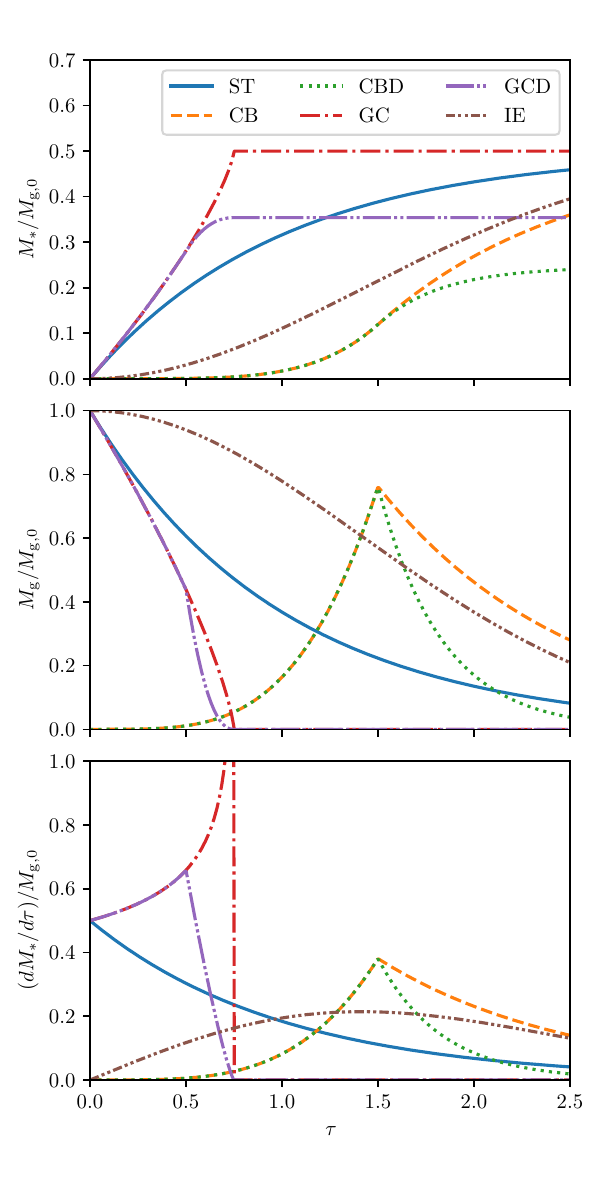}
\caption{
\label{fig:example_models}
Example evolutionary histories of stellar mass (top), gas mass (middle), and star formation rate (bottom) for each of the models discussed in the paper (as indicated in the legend). For the purposes of this plot, we use $\eta = 1$ in all models. For CB and CBD we use $\taua = 1.5$ and $p=3$, for CBD {and GCD} we use $\eta_{\rm d} = 5$, for GC we use $\xi=1$ and $\tauc = 0.75$, {for GCD we use $\xi=1$, $\tauc = 0.75$, and $\taufb = 0.5$,} and for IE we use $\chi=0.5$ and $\delta=1$. See main text for definitions of the various parameters.
}
\end{figure}

Having established this general framework, we now consider a range of scenarios for how a star cluster might be assembled. We plot example histories for each model in \autoref{fig:example_models}, and summarise the models and their key free parameters in \autoref{tab:model_list}.

\subsubsection{Static cloud (ST)}
\label{sssec:pre-assembled}

Our first scenario is the simplest possible, a non-accreting cloud with constant $\tff$ that begins with an initial gas mass $\Mg = \Mgz$ and starts forming stars at time $t=0$. We refer to this as the static, or ST model, hereafter. {Of course, if the density and free-fall time are constant, but the gas mass is not, then this means that the cloud is not static in terms of its radius; since the data to which we will compare below do not include detailed information on the spatial structure of stellar populations, however, the constant free-fall time is the property that is relevant for our purposes.} Physically, this would correspond to a situation where cloud assembly is rapid compared to the process of star formation, or where a cloud is assembled in a state where it cannot form stars immediately. As first pointed out by \citet{ginsburg12a} and \citet[also see \citealt{walker16a} and \citealt{urquhart18a}]{longmore14a}, such a scenario can be ruled out for at least the most massive young clusters observed in the Milky Way, on the grounds that there are no observed gas clouds dense and massive enough to be the progenitors of the most massive clusters. On the other hand, \citet{krumholz19a} point out that there is no such difficulty for clusters near the Galactic Centre, and in this region there do indeed appear to be very massive and dense molecular clouds with little or no star formation activity such as ``the Brick'' \citep{longmore13a, rathborne14b}. These have been hypothesised to remain quiescent until star formation within them is triggered by a pericentre passage around Sgr A$^*$ \citep{kruijssen15a}, and thus they represent potential exemplars of the static cloud scenario{, though recent observations of infall in at least some of these objects suggest something closer to one of the alternative scenarios we describe below \citep{barnes19a}}.

Since there is no mass accretion in this model, $\dMa=0$, and we will also assume $\eta$ is constant, the solution to \autoref{eq:evol_eq} is trivial:
\begin{equation}
\label{eq:ms_st}
\Ms = \frac{\Mgz}{1+\eta}\left(1-e^{-\tau}\right)
\qquad
\Mg = \Mgz e^{-\tau},
\end{equation}
where $\tau = t/\tsf$ and
\begin{equation}
\label{eq:tsf}
\tsf = \frac{\tff}{(1+\eta) \eff}
\end{equation}
is the star formation timescale; this is the natural timescale over which the star formation process occurs, and the cloud is converted to stars or dispersed. The final star formation efficiency (SFE), defined as the ratio of final stellar mass to total mass of gas available for star formation, is
\begin{equation}
\epsilon_* \equiv\frac{M_*}{\Mgz}= \frac{1}{1+\eta}.
\end{equation}

\subsubsection{Conveyor belt (CB)}
\label{sssec:conveyor}

The absence of gas clouds as massive and dense as the densest star clusters, as noted in \autoref{sssec:pre-assembled}, led \citet{longmore14a} to propose a ``conveyor belt'' model where gas accretion occurs simultaneously with cluster formation, so that the full mass of the gas cloud is never assembled at a single time; observations that regions such as the ONC frequently sit at the intersections of filaments supports this picture \citep{motte18a}. {In this picture, stars may form in both the filaments and in the central hub, but stars that wind up as part of a bound cluster at the end of the star formation process are mostly those that form in the central hub. This hub region is continually re-supplied by accretion of gas from the filaments. For the purposes of this paper, and for the data sets to which we will compare below, we are primarily interested in what happens in the hub.}

In principle the region fed by a conveyor belt could be static, expanding, or contracting. Numerical simulations and analytic calculations by a number of authors \citep[e.g.,][]{klessen00a, goldbaum11a, matzner15a, lee16b, lee16c} suggest that, as long as {the accretion rate is high enough that a cloud's growth time is comparable to its free-fall time, the inflow supplies enough energy for the density and star formation rate per free-fall time to remain roughly constant for multiple free-fall times. Examples of such models include \citet[their Figure 3]{goldbaum11a}, \citet[the $10^5$ and $10^6$ $M_\odot$ models shown in their Figure 1]{zamora-aviles14a}, and \citet[their Figure 6]{lee16c}: in all of these models, the free-fall time varies by no more than a factor of $\sim 2$ over multiple cloud free-fall times. For this reason} we will assume constant $\tff$ and $\eff$. We refer to this model as conveyor belt, or CB, hereafter.

We abstract this model as having an initial gas mass of zero, and an accretion rate that varies in time as a powerlaw $t^p$. We generically expect $p > 0$, since gravity-driven accretion rates generally rise with time until the reservoir of mass is exhausted; \citet{goldbaum11a} show that pressureless collapse of a reservoir of constant surface density that becomes gravitationally unstable naturally produces $p \approx 3$; \citet{lee16c} find a similar value of $p$ while protoclusters are small compared to their parent reservoirs, but that this tapers to $p\approx 0$ once $\gtrsim 10\%$ of the parent reservoir has been accreted. We will adopt the \citeauthor{goldbaum11a} value of $p=3$ as our fiducial choice, but for completeness we give the model result for general $p$, by taking the accretion rate to be
\begin{equation}
\dMa = H(\ta-t) (p+1) \frac{\Mgz}{\ta} \left(\frac{t}{\ta}\right)^p,
\end{equation}
where $\Mgz$ is the total mass that will eventually reach the protocluster, $\ta$ is the time over which accretion happens, and $H(x)$ is the Heaviside step function. The initial conditions are $\Mg = M_* = 0$. With this accretion rate, \autoref{eq:evol_eq} has the following analytic solutions for any non-negative integer $p$:
\begin{eqnarray}
\label{eq:ms_cb}
M_* & = & 
\left\{
\begin{array}{ll}\displaystyle
\frac{\Mgz}{(1+\eta)(p+2)\taua^{p+1}} g(\tau,p+2),
& \tau \leq \taua 
\vspace{0.2cm}\\
\displaystyle
M_*(\taua) + \frac{\Mg(\taua)}{1+\eta} \left(1 - e^{-\tau+\taua}\right), & \tau > \taua
\end{array}
\right. \\
\label{eq:mg_cb}
M_g & = & 
\left\{
\begin{array}{ll}\displaystyle
\frac{\Mgz}{\taua^{p+1}} g(\tau,p+1), & \tau \leq \taua 
\vspace{0.2cm}\\ \displaystyle
\Mg(\taua) e^{-\tau+\taua}, & \tau > \taua
\end{array}
\right.
\end{eqnarray}
where for $p\geq 1$
\begin{eqnarray}
g(\tau,p) &=& p e^{-\tau}\int_0^\tau\tau'^{(p-1)} e^{\tau'} d\tau',
\label{eq:gdef}\\
&=& p! \left[ (-1)^p e^{-\tau} - \sum_{i=1}^p \frac{(-1)^i}{(p-i)!} \tau^{p-i}\right].
\end{eqnarray}
Here $\tau = t/\tsf$ as in \autoref{sssec:pre-assembled}, $\taua = \ta/\tsf$,
and we made use of the relations
\begin{eqnarray}
\int_0^\tau g(\tau',p)d\tau'& =& \frac{g(\tau,p+1)}{p+1},
\label{eq:intg}\\
&=&\tau^p-g(\tau,p).
\label{eq:intg2}
\end{eqnarray}
To get a feeling for the magnitude of $g(\tau,p)$, we note that $g(\tau,1)=1-e^{-\tau}$ and that $g(\tau,2)=2(\tau-1+e^{-\tau})$. {The approximation}
\begin{equation}
{
g(\tau,p)\simeq \frac{\tau^{p}}{1 + \tau/p}
}
\end{equation}
{is accurate to better than 15\%.} Next, observe that \autoref{eq:intg} implies 
\begin{equation}
\frac{dg(\tau,p+1)}{d\tau}=(p+1)g(\tau,p).
\label{eq:dg}
\end{equation}
In turn, this relation implies that $g(\tau,p)$ is a monotonically increasing function of $\tau$
since \autoref{eq:gdef} implies that $g(\tau,p)$ is positive. It follows from \autoref{eq:mg_cb} that the gas mass increases monotonically until the accretion stops.

At times $\tau \gg \taua$, the star formation efficiency in the conveyor belt model approaches $\epsilon_* = 1/(1+\eta)$, exactly as in the static cloud case, but the star formation history is different. This model satisfies the observational constraint that originally motivated it, in that the gas mass need never be large compared to the final stellar mass. Indeed, the final stellar mass (achieved in the limit $\tau\to\infty$) is $M_{*,\rm f} = \Mgz/(1+\eta)$ regardless of the accretion history, while the maximum gas mass (achieved when $\tau=\taua$) is {$M_{\rm g,max} \approx \Mgz/[1+\taua/(p+1)]$}. Thus as long as $\taua \gtrsim \eta$, the maximum gas mass will be comparable to or smaller than the final stellar mass.

An important feature of this conveyor belt model is that {\it star formation always accelerates while gas is accreting}, provided $p\geq 0$. With the aid
of \autoref{eq:dg}, we find that the acceleration in the stellar mass is
\begin{equation}
\ddot M_*=\frac{p+1}{(1+\eta)\taua^{p+1}}\left(\frac{\Mgz}{\tsf^2}\right) g(\tau,p+2)~~~~~~~\tau\leq\taua,
\end{equation}
which is always positive, as noted above. Such acceleration appears to be demanded by the observations \citep{palla00a}.

\subsubsection{Conveyor belt plus rapid dispersal (CBD)}
\label{sssec:conveyor_plus}

A slight variation on the standard conveyor belt model is to note that, as pointed out by \citet{goldbaum11a}, mass loss rates are likely sensitive to the strength of the confining ram pressure from accretion. Consequently, it makes sense to adopt a mass loading factor that increases significantly once accretion ceases, leading to more rapid dispersal. We refer to conveyor belt models in which dispersal after the end of accretion is rapid as conveyor belt plus dispersal (CBD) models hereafter. From the standpoint of our simple analytic models, we can model this by setting $\eta$ to one value during the accretion phase, $t < \ta$, and to another value $\eta_{\rm d} > \eta$ during the dispersal phase, $t > \ta$. In all other respects this model is identical to the simple conveyor belt model of \autoref{sssec:conveyor}. The solution to \autoref{eq:evol_eq} in this case is modified only slightly from that given by \autoref{eq:ms_cb} and \autoref{eq:mg_cb}:
\begin{eqnarray}
\label{eq:ms_cbd}
M_* & = & 
\left\{
\begin{array}{ll}\dis
\frac{\Mgz}{(1+\eta)(p+2)\taua^{p+1}} g(\tau,p+2), & \tau \leq \taua 
\vspace{0.2cm}\\ \dis
M_*(\taua) + \frac{\Mg(\taua)}{1+\eta_{\rm d}} \left(1 - e^{-\phid(\tau-\taua)}\right) & \tau > \taua
\end{array}
\right. \\
\label{eq:mg_cbd}
M_g & = & 
\left\{
\begin{array}{ll}\dis
\frac{\Mgz}{\taua^{p+1}} g(\tau,p+1), & \tau \leq \taua 
\vspace{0.2cm}\\
\Mg(\taua) e^{-\phid(\tau-\taua)} & \tau > \taua
\end{array}
\right.
\end{eqnarray}
where
\begin{equation}
\phid \equiv \frac{1 + \eta_{\rm d}}{1 + \eta}
\label{eq:phid}
\end{equation}
can be thought of as representing the ratio of star formation efficiencies during and after the accretion phase. This model shares the key feature of the conveyor belt model: there is no need to assemble a cloud as massive as the final star cluster all at once, since the histories are identical up to the end of the accretion phase, but then a smaller fraction of the remaining gas mass is converted to stars than in the standard conveyor belt case. To be precise, the final star formation efficiency is
\begin{equation}
\epsilon_* = \frac{1}{1+\eta}
\left[1 - \left(\frac{\phid-1}{\phid}\right) \frac{g(\taua,p+1)}{\taua^{p+1}}\right]
\label{eq:epsstar_cbd}
\end{equation}
Equations (\ref{eq:intg}) and (\ref{eq:intg2}) imply that the ratio $g(\tau,p+1)/\tau^{p+1}$ is strictly smaller than unity for any $\tau>0$ since $g(\tau,p+2)>0$, so the final star formation efficiency is between $1/(1+\eta)$ and $1/(1+\eta_{\rm d})$.

\subsubsection{Global collapse (GC)}
\label{sssec:ghc}

The observation that star formation accelerates could be a reflection of gas accumulation, as in the CB or CBD models, but it could also be a result of the star formation process itself. An example of such a model is the global collapse (GC) scenario proposed by a number of authors (e.g., \citealt{zamora-aviles14a, kuznetsova15a, kuznetsova18a, vazquez-semadeni17a, vazquez-semadeni19a}). The central idea of GC models is that clouds are assembled in a low density state but then undergo a global collapse. Consequently, the mean free-fall time, rather than remaining constant, systematically decreases on a free-fall timescale as the mean density rises. The combination of an apparently-extended star formation history and an accelerating star formation rate is then taken to be due to the decreasing free-fall time: stars that form at early times may have ages comparable to the free-fall time of the system when the formed, but this can be significantly longer than the free-fall time of the system at the time when it is observed. Moreover, as the system gets denser, the free-fall time decreases and thus star formation accelerates.

{In terms of the hub-and-filament geometry frequently observed in star-forming regions, and discussed in \autoref{sssec:conveyor}, the difference between the CB (or CBD) and GC models is the assumed time evolution of the hubs. In the CB model, the hub is assumed to remain at roughly constant density over many free-fall times, so that any acceleration of star formation is due to the mass of the hub increasing, not due to its density rising. By contrast, in GC the hub is assumed to be in a process of collapse on a dynamical timescale (even if it is also accreting), so that the density rises with time, and this accounts for most or all of the increase star formation rate with time. Examples of published models in the latter category include the $10^3$ or $10^4$ $M_\odot$ cases shown in Figure 1 of \citet{zamora-aviles14a}, where, once the clouds grow massive enough, the density runs away to infinity on roughly a free-fall timescale.}

Mathematically we can represent this model by assuming that the mean density obeys
\begin{equation}
\frac{d\rho}{dt} = \xi \frac{\rho}{\tff(\rho)}
\end{equation}
where $\tff(\rho) = \sqrt{3\pi/32 G \rho}$ is the free-fall time at the current density. The constant $\xi$ specifies how fast the cloud contracts compared to the free-fall timescale, with higher $\xi$ corresponding to more rapid contraction. The value of $\xi$ will depend at least partly on geometry -- $\xi \approx 1$ is expected for 3D structures, but values as small as $\sim 0.1$ are possible for highly-flattened geometries {if one interprets $\rho$ as the density internal to the structure} \citep{toala12a}{; however, note that if one interprets $\rho$ as the mean density of a spherical structure of the same size, as is frequently done when interpreting observations, then $\xi \gtrsim 1$ even for flattened structures}. For a cloud that starts at density $\rho_0$ at time $t=0$, the density and free-fall time evolve as
\begin{equation}
\label{eq:tff_ghc}
\rho = {\frac{\rho_0}{\left(1 - x\right)^2}}
\qquad
\tff = {\tffz \left(1 - x\right)}
\end{equation}
where {$x = t/\tc$, $\tc = 2 \tffz/\xi$ is the time at which the cloud reaches infinite density, and $\tffz = \sqrt{3\pi/32 G \rho_0}$ is the initial free-fall time.}

Inserting this non-constant free-fall time into \autoref{eq:evol_eq}, holding $\eta$ and $\eff$ constant, and solving subject to the initial condition that $\Mg = \Mgz$ and $M_* = 0$ at $t=0$, we obtain
\begin{eqnarray}
\label{eq:ms_ghc}
M_* & = & 
\frac{\Mgz}{1+\eta}
\left\{
\begin{array}{ll}
1 - \left(1 - {x}\right)^{\tauc},
& {x<1} \\
1, & {x \geq 1}
\end{array}
\right.
\\
\label{eq:mg_ghc}
\Mg & = & \Mgz
\left\{
\begin{array}{ll}
\left(1 - {x}\right)^{\tauc},
& {x<1} \\
0, & {x\geq 1}
\end{array}
\right.,
\end{eqnarray}
The quantity
\begin{equation}
\label{eq:tauc_defn}
\tauc =  \frac{2 (1+\eta)\eff}{\xi}
\end{equation}
is the dimensionless time at which the cloud collapses to infinite density and $\tff \to 0$, {where we have non-dimensionalised time using} $\tau = t/\tsf$ as before, but we now define $\tsf = \tffz/[(1+\eta)\eff]$ (c.f.~\autoref{eq:tsf}), i.e., we define $\tsf$ using the initial free-fall time since $\tff$ is non-constant. Half the stars have formed and half the gas has been consumed at a time
\beq
t_{1/2}=\left(1-\frac{1}{2^{1/\tauc}}\right)\tc,
\eeq
and correspondingly the free-fall time then is
\beq
t_{\rm ff,\,1/2}=\frac{\tffz}{2^{1/\tauc}}.
\eeq
For $\tauc\ga 1$, half the stars form at a rate not that different from the initial rate. {Indeed, in the limit $\xi \ll 1$, and thus $\tauc\gg 1$, the GC model approaches the ST model, since the collapse then becomes slow compared to star formation. (Conversely, in the limit $\eff \to 1$, the ST and CB models become qualitatively similar to GC, since then all gas is converted to stars on a dynamical timescale.)} More generally, the rate at which the star formation rate changes is
\begin{equation}
\ddot{M}_* = \frac{\Mgz}{(1+\eta)\tsf^2}\left(\frac{1-\tauc}{\tauc}\right) \left(1 - \frac{\tau}{\tauc}\right)^{\tauc-2},
\end{equation}
so star formation accelerates with time ($\ddot{M}_* > 0$) only if $\tauc < 1$. The final star formation efficiency is $\epsilon_* = 1/(1+\eta)$, exactly as in the ST or CB models. 

\subsubsection{{Global collapse plus dispersal (GCD)}}

{Just as the CBD model adds a more rapid dispersal phase (i.e., a larger value of $\eta$) to CB, one can similarly posit a GC model with rapid dispersal at its end. In the CBD model the natural cause of an increase is the removal of confinement by the accretion flow. In GC there is no similar natural breakpoint, but a number of authors \citep[e.g.,][and references therein]{vazquez-semadeni19a} have posited that the stellar initial mass function (IMF) is time-dependent, so that massive stars only form late in the collapse process. If this hypothesis were correct, it would naturally cause the mass loading factor to increase at later times. Mathematically, we model this by introducing two new free parameters: $\phid$, which is defined exactly as for the CBD model (\autoref{eq:phid}) as the ratio of star formation efficiencies before and after massive star feedback ``turns on'', and $\tfb$, which represents the time at which this happens.}

{
If we let $\eta$ be the mass loading parameter prior to $t < \tfb$, $\eta_{\rm d} = \phid(1+\eta)-1$ be the mass loading factor from from $\tfb < t < \tc$, and continue to use \autoref{eq:tff_ghc} to describe the evolution of the free-fall time, the solution to \autoref{eq:evol_eq} is
}
\begin{eqnarray}
\lefteqn{
{
M_* = \frac{\Mgz}{1+\eta}\cdot{}
}
}
\nonumber \\
& &
{
\left\{
\begin{array}{ll}
1- \left(1 - x\right)^{\tauc}, & x < x_{\rm fb} \\[1.5ex]
\phid^{-1} \left(1-x_{\rm fb}\right)^{\tauc} \left[1 - \left(\frac{1-x}{1-x_{\rm fb}}\right)^{\phid \tauc} \right]
+ {} \\
\qquad 1 - \left(1-x_{\rm fb}\right)^{\tauc},
& x_{\rm fb} \leq x < 1 \\[1.5ex]
1 - \left(\frac{\phid-1}{\phid}\right)
\left(1-x_{\rm fb}\right)^{\tauc},
& x \geq 1
\end{array}
\right.
}
\label{eq:ms_gcd}
\\
\lefteqn{
{
\Mg = \Mgz \cdot{}
}
}
\nonumber \\
& & 
{
\left\{
\begin{array}{ll}
\left(1-x\right)^{\tauc}, & x < x_{\rm fb} \\[1.5ex]
\left(1 - x_{\rm fb}\right)^{\tauc} \left(\frac{1-x}{1-x_{\rm fb}}\right)^{\phid\tauc}, & x_{\rm fb} \leq x < 1 \\[1.5ex]
0, & x \geq 1
\end{array}
\right.
}
\label{eq:mg_gcd}
\end{eqnarray}
{
where $x_{\rm fb} = \tfb/\tc$. The final star formation efficiency is
}
\begin{equation}
{
\epsilon_* = \frac{1}{1+\eta}
\left[1 - \left(\frac{\phid-1}{\phid}\right)\left(1-x_{\rm fb}\right)^{\tauc}\right].
}
\label{eq:epsstar_gcd}
\end{equation}
{
As with CBD (c.f.~\autoref{eq:epsstar_cbd}), the factor inside the square brackets is strictly negative, and thus the final star formation efficiency is lower than in the corresponding model without the disruption phase. Star formation continues accelerating during the gas clearing phase only if $\phid\tauc < 1$; otherwise it decelerates.
}

\subsubsection{Increasing star formation efficiency (IE)}
\label{sssec:accel}

A final potential mechanism to explain why star formation accelerates in protoclusters like the ONC is to posit that this is an intrinsic part of the star formation process itself. \citet{lee15b} and \citet{murray15a} argue that, rather than being constant, $\eff$ increases with time in star-forming regions as $\eff \propto t^\delta$, with $\delta \approx 1$; we refer to this as the increasing efficiency (IE) model. Although somewhat similar to the GC model, the two are conceptually distinct in that star formation accelerates in the GC model because the mean density rises with time, while in the IE model it accelerates even though the mean density remains constant because the star formation process itself becomes more efficient. Mathematically, the two models differ in their predicted rate of acceleration. \citet{caldwell18a} argue that the IE model provides a good fit to observed star formation histories in resolved clusters, and \citet{lee16a} and \citet{ochsendorf17a} argue it provides a good fit to the observed ratio of ionising luminosity to CO luminosity, though, as we note above, the quality of the agreement is extremely sensitive to the choice of procedure for matching up non-co-spatial molecular gas and H~\textsc{ii} regions.

For the purposes of comparing this model to data, we adopt the same parameterisation as \citet{lee16a}: $\eff = \effz (t/\tff)^\delta$. Thus $\effz$ represents the value of $\eff$ one free-fall time after the onset of star formation. While the theoretical models of \citet{lee15b} and \citet{murray15a} give $\delta = 1$, we will allow $\delta$ to be a free parameter from $0-3$ when we fit to observations below. The solution to \autoref{eq:evol_eq} for arbitrary $\delta \geq 0$, holding $\eta$ and $\tff$ constant, subject to the initial conditions $\Mg=\Mgz$ and $M_* = 0$ at $t=0$, is
\begin{eqnarray}
\label{eq:ms_ie}
M_* & = & \frac{\Mgz}{1+\eta}\left[1 - \exp\left(-\frac{\chi^\delta \tau^{1+\delta}}{1+\delta}\right)\right] \\
M_g & = & \Mgz \exp\left(-\frac{\chi^\delta \tau^{1+\delta}}{1+\delta}\right) 
\label{eq:mg_ie}
\end{eqnarray}
where $\tau = t/\tsf$, $\tsf = \tff/[(1+\eta)\effz]$ (i.e., we define $\tsf$ using the value of $\eff$ at 1 free-fall time; c.f.~\autoref{eq:tsf}), and $\chi = \tsf/\tff
=1/[(1+\eta)\effz]
$. The final star formation efficiency is $\epsilon_* = 1/(1+\eta)$, exactly as in the static model. The average efficiency with which stars form is
\begin{eqnarray}
\overline{\epsilon}_{\mathrm{ff}}& = & \frac{1+\eta}{\Mgz} \int_0^{\infty} \eff(\tau) \dot{M}_*(\tau)\, d\tau
\nonumber
\\
& = & 
\left[\chi(1+\delta)\right]^{\delta/(1+\delta)}\Gamma\left(1+\frac{\delta}{1+\delta}\right) \effz.
\label{eq:eff_mean}
\end{eqnarray}
For typical parameters in this model, $\delta = 1$ and $\chi = 50$, this gives $\overline{\epsilon}_{\rm ff} \approx 8.9 \effz$, so most stars form at an efficiency substantially higher than that which prevails for the first free-fall time. Intuitively, this makes sense: in this model there are a relatively long period of near-quiescence when $\eff$ is small and few stars form, but this is followed by a burst of activity after $\eff$ becomes large; most stars form during this final burst. Quantitatively, the second derivative of the star formation rate is
\begin{equation}
\ddot{M}_* = \frac{\Mgz}{(1+\eta) \tsf^2} \left[\chi^\delta \tau^{\delta-1} \left(\delta - \chi^\delta \tau^{\delta + 1}\right) \exp\left(-\frac{\chi^\delta \tau^{1+\delta}}{1+\delta}\right)\right].
\end{equation}
The sign of $\ddot{M}_*$ therefore depends on $\delta - \chi^\delta \tau^{\delta+1}$; for sufficiently small $\tau$ this term is positive, and star formation accelerates. Later on, as gas is depleted, this term becomes negative and star formation decelerates.


\section{Confrontation with observations}
\label{sec:obs}

Having outlined the various models, we now compare them to observations.

\subsection{Star formation histories}
\label{ssec:sf_hist}

\subsubsection{Data set}
\label{sssec:data_set}

The first observation to which we are interested in comparing is the observed distribution of stellar ages in young clusters; as discussed in \autoref{sec:intro}, working through the implications of the observed extended but accelerating star formation histories in such regions is one of our primary motivations in this work. For our observational data set, we select two young open clusters: the Orion Nebula Cluster and NGC 6530. We focus on these two because they both offer very clean, high-quality data: membership lists determined from \textit{Gaia} 6D phase space data plus other ancillary indicators, and ages determined from spectroscopy, with star-by-star extinction corrections. The free-fall time in the ONC is $t_{\rm ff,ONC} \approx 0.6$ Myr as determined from dynamical modelling by \citet{da-rio14a}. For NGC 6530, \citet{prisinzano19a} measure a stellar velocity dispersion of $\sigma_{\rm NGC6530} = 2.42$ km s$^{-1}$, and the effective radius of the cluster is $0.1^\circ$ \citep{kharchenko13a}, which translates to $2.3$ pc for the best-fit distance of 1.32 kpc obtained by \citeauthor{prisinzano19a}. Thus the crossing time is $t_{\rm cr} = r_{\rm NGC6530}/\sigma_{\rm NGC6530} = 0.93$ Myr. For a virialised object, the free-fall time is approximately half the crossing time \citep{tan06a}, so we adopt $t_{\rm ff,NGC6530} = 0.5$ Myr.

For our stellar ages in NGC 6530, we use the fits provided by \citet{prisinzano19a}. For the ONC, we must select down from the full catalog of \citet{kounkel18a}, since their study covers the entire Orion star-forming region and includes multiple populations across a large volume. For this study, we select stars from their catalog that are within 1 pc in projection of $\theta^1$ C (the same radius within which we have estimated the free-fall time), and that are kinematically identified as part of the Orion A population. We take the ages of these stars from \citeauthor{kounkel18a}, using only the ages based on spectroscopic determinations, since those based on colour are unreliable in the ONC due to high extinction. After applying these cuts, our sample consists of 185 stars in the ONC and 395 stars in NGC 6530.

In addition to the age estimates themselves, in order to carry out a meaningful statistical analysis we must have some understanding of the uncertainties in the measurements. Uncertainties in the ages of young stars has been a topic of considerable debate in the literature in recent years, and we refer to the readers to the reviews by \citet{soderblom14a}, \citet{jeffries17a}, and \citet{krumholz19a} for a detailed discussion. Young stellar ages are always subject to a systematic uncertainty of $\sim 0.1-0.3$ dex in the \textit{absolute} age scale coming from the choice of pre-main sequence tracks. However, there is significantly less uncertainty in the \textit{relative} ages of stars \citep[e.g.][]{reggiani11a}, which is the quantity of concern for us, since we are interested in the star formation history -- a shift in absolute age just amounts to a rescaling of the timescales.

Relative age uncertainties come from a variety of factors, depending on the age-dating method. Uncertainties larger than $\approx 0.2-0.3$ dex can be ruled out by independent methods of constraining dispersions of stellar age (e.g.,~radii derived from rotation or gravity-sensitive spectral features -- \citealt{jeffries07a, da-rio16a, prisinzano19a}), but a range of estimates below this limit have been published \citep[e.g.,][]{preibisch12a, da-rio16a, prisinzano19a}. For this work we adopt the results of \citet{prisinzano19a}: we take the error in log age to be a Gaussian with a width $\sigma = 0.13$ dex and a bias $b = -0.05$ dex (i.e., true stellar ages are on average 0.05 dex older than estimated ones). The systematic bias is due to unresolved binarity, which increases luminosity at fixed effective temperature, and thus tends to bias age estimates low. We have experimented with other choices of these parameters, subject to the overall constraint that the total error cannot exceed $\approx 0.2-0.3$ dex, and we find that the posterior PDFs for some parameters can be sensitive to the exact choice of $\sigma$ and $b$, as are quantitative measures of relative goodness-of-fit such as the Akaike information criterion. Since we do not understand the true error distribution in detail, we will for this reason limit our analysis to general features that are robust against plausible changes in $\sigma$ or $b$.

\subsubsection{Likelihood function}

\begin{table}
\caption{
\label{tab:definitions}
Definitions of parameters used in computing the stellar age distribution likelihood function.
}
\begin{tabular}{ll}
\hline\hline
Parameter & Meaning \\ \hline
$\tobs$ & Age of cluster (time since onset of star formation) \\
$t_*$ & True age of a star \\
$t_{*,\rm obs}$ & Observationally-estimated stellar age (including errors) \\
$\sigma$ & Dispersion of stellar age error distribution \\
$b$ & Bias in the stellar age error distribution \\
$t_{\rm ff,clust}$ & Present-day free-fall time in cluster \\
$\epsilon_{\rm *,clust}$ & Present-day star formation efficiency, $M_*(\tobs)/\Mgz$ \\
$f_{\rm g,clust}$ & Present-day gas fraction, $\Mg/ (\Mg + \Ms)$ at $t=\tobs$ \\
\hline\hline
\end{tabular}
\end{table}

We wish to compare the observed age distribution to that predicted by our various candidate models. To this end, we now compute a likelihood function, which gives the probability density of the data given the model. For convenience we summarise the meanings of various parameters that we introduce in this calculation in \autoref{tab:definitions}.

For a cluster formation model with stellar mass as a function of dimensionless time, $M_*(\tau)$, the distribution of log stellar ages that will be seen a time when the cluster age is $\tobs$ (i.e., a time $\tobs$ after the onset of star formation) is 
\begin{eqnarray}
\frac{dp}{d\log t_*} & = &  \left(\ln 10\right) \frac{t _* M'_*(\tauobs - \tau_*)}{\tsf M_*(\tauobs)}
\nonumber  \\
& = &\left(\ln 10\right) \eff \left(\frac{t_*}{\tff}\right)\frac{M_g(\tauobs - \tau_*)}{M_*(\tauobs)},
\label{eq:dpdlogt}
\end{eqnarray}
where $t_*$ is the stellar age, $\tauobs = \tobs/\tsf$ and $\tau_* = t_*/\tsf$ are the dimensionless cluster and stellar ages, respectively, and $M_*' = dM_*/d\tau$. The factor of $\ln 10$ is to ensure that the PDF is properly normalised to have unit integral over all $\log t_*$. The stellar mass versus dimensionless time, $M_*(\tau)$, is given by \autoref{eq:ms_st}, \autoref{eq:ms_cb}, \autoref{eq:ms_cbd}, \autoref{eq:ms_ghc}, {\autoref{eq:ms_gcd},} and \autoref{eq:ms_ie}, for the ST, CB, CBD, GC, {GCD,} and IE models, respectively; the corresponding gas masses, $M_g(\tau)$, are given by \autoref{eq:ms_st}, \autoref{eq:mg_cb}, \autoref{eq:mg_cbd}, \autoref{eq:mg_ghc}, {\autoref{eq:mg_gcd},} and \autoref{eq:mg_ie}.

Note that, in the GC {and GCD} model{s}, $\tff$ is also a function of $\tauobs-\tau_*$ (\autoref{eq:tff_ghc}). {These models produce} a double-peaked profile in the distribution $dp/d\log t_*$; \autoref{eq:dpdlogt} shows that the age distribution is proportional to $\tau_* M'_* (\tauobs - \tau_*)$, or, in terms of the parameter $\tau$ in \autoref{fig:example_models}, $(\tauobs - \tau) M'_*(\tau)$. Reference to \autoref{fig:example_models} shows that this leads to a double peak in the GC {and GCD} model{s}, with one peak at $\tau\sim \tauobs$ and a second at $\tau\sim\tauobs-\tauc$ {or $\tau\sim\tauobs-\taufb$}. 

To incorporate the effects of errors, we convolve the true age distribution with the error distribution. Following our discussion in \autoref{sssec:data_set}, we parameterise the uncertainty distribution in log age as a biased Gaussian, i.e., for a star whose \textit{true} log age is $\log t_*$, the distribution of \textit{measured} log ages $\log t_{\rm *,obs}$ is
\begin{equation}
f(\log t_{\rm *,obs} \mid \log t_*) =
\frac{1}{\sqrt{2\pi\sigma^2}} \exp\left[-\frac{\left(\log t_* - \log t_{\rm *,obs} + b\right)^2}{2\sigma^2}\right],
\end{equation}
where $b$ is the bias and $\sigma$ is the dispersion, and both $b$ and $\sigma$ are in units of dex. The full distribution of observed ages is therefore given by
\begin{equation}
\label{eq:dmtobs}
\frac{dp}{d\log t_{\rm *,obs}} = \int_{-\infty}^\infty \left( \frac{dp}{d\log t_*}\right) f(\log t_{\rm *,obs} \mid \log t_*) \, d\log t_*.
\end{equation}
We evaluate this integral numerically via Fourier transform, since it is equivalent to the convolution of the true stellar age distribution $dp/d\log t_*$ with a Gaussian. The log likelihood function $\mathcal{L}$ is simply the probability density of the data given the model:
\begin{equation}
\log \mathcal{L} = \sum_{i=1}^N \log \left(\frac{dp_*}{d\log t_{\rm *,obs}}\right)_{t_{\rm *,obs}=t_i}
\end{equation}
where $t_i$ is the age estimated for the $i$th star in the observed sample.

Our stellar age distributions as written depend on two dimensional quantities: the cluster age $\tobs$, and the star formation timescale $\tsf$ that scales between physical times $t$ and dimensionless times $\tau=t/\tsf$. We treat these as free parameters to be fit. In addition, we fit free parameters for each of the models: $\ta$ for model CB, $\ta$ and $\phid$ for model CBD, $\tc$ for model GC, {$\tc$, $\tfb$, and $\phid$ for model GCD,} and $\delta$ for model IE. Note that we do not have to fit to $\eta$ or $\xi$ (for the GC {and GCD} model{s}), because $\eta$ is absorbed into the definition of $\tsf$, and $\xi$ into the definition of $\tc$. We adopt priors that are flat in the logarithm of all the positive-definite quantities (all timescales) or that are strictly greater than unity ($\phid$), and flat linear priors in all other parameters. We impose almost no constraint on the time of observation $\tobs$, allowing any value in the range $0.01 - 100$ Myr, but we limit the allowed ranges of the remaining parameters based on physical considerations, which we now proceed to describe.

First, for all models we set the prior probability to zero for $\eff$ outside the range $10^{-4}$ to 1, on the grounds that $\eff$ values outside this range correspond to unphysically-inefficient or efficient star formation; to estimate $\eff$ from $\tsf$, we use the observed free-fall time in NGC 6530 or the ONC, as appropriate, and $\eta = 1$.\footnote{Applying this prior to the GC {and GCD} case{s} requires some care, because a particular combination of $\tobs$, $\tsf$, and $\tc$, the parameters to which we are fitting, does not by itself determine a unique value of $\eff$; instead, one can change $\eff$ arbitrarily while leaving all these timescales unchanged by simultaneously changing $\xi$ and $\tffz$. To determine $\eff$, we must therefore choose a value of $\xi$. We can do so by considering two possible scenarios. One is that the cluster in question has not yet reached collapse ($\tobs < \tc$), in which case we can fix $\xi$ by demanding that the free-fall time in the model match the observed present-day free-fall time $t_{\rm ff,clust}$ (0.6 Myr for the ONC, 0.5 Myr for NGC 6530, respectively). Re-arranging \autoref{eq:tff_ghc}, we find that the value of $\xi$ that satisfies this condition is $\xi = 2 t_{\rm ff,clust}/(\tc-\tobs)$. This in turn breaks the degeneracy and allows us to determine a unique value of $\eff$. The other possibility is that the cluster as we see it today is after the collapse to singularity ($\tobs > \tc$), in which case the free-fall time we measure is a result of the stars rebounding to their current positions post-collapse, and has nothing to do with the free-fall time prior to collapse. In this case $\xi$ is unconstrained by the fit, and we must therefore adopt a value of $\xi$. For this case we choose a fiducial value $\xi = 1$. Our calculation of the best-fitting model is able to consider both scenarios, since we do not impose any prior on whether $\tobs < \tc$ or $\tobs > \tc$. \label{note:ghc_xi}} This serves to define the allowed range of $\tsf$. Second, we apply priors based on the physical picture that motivates each model. For the CB and CBD models, the physical picture is that accretion is due to the collapse of a larger-scale, lower-density reservoir with a longer dynamical time than the cluster-forming region, a picture that requires $\ta > \tff$; we also require $\ta \leq \tobs$, not for any physical reason, but simply because all models with $\ta > \tobs$ have identical age distributions for the stars that exist today, and thus cannot be distinguished in our analysis. For the GC {and GCD} model{s}, the central idea is that regions collapse on a free-fall timescale, forming stars while doing so. We therefore impose as a prior $0.1 < \xi < 10$; lower values of $\xi$ correspond to collapses so slow as to be nearly indistinguishable from the ST model, while higher values require regions to collapse much faster than a free-fall time, which is unphysical. This serves to limit the range of $\tc$ (see \autoref{note:ghc_xi}). Finally, for IE, theoretical models of how the density structure changes as star formation proceeds predict $\delta \approx 1$. We allow some range around this, by setting our prior to zero outside the range $\delta = 0 - 3$.

Our third and final prior is on the present-day star formation efficiency, $\epsilon_{*,\rm clust} \equiv M_*(\tobs) / \Mgz$, i.e., the fraction of all the gas available that has been converted to stars; note that $\epsilon_{*,\rm clust}$ may be smaller than the final star formation efficiency $\epsilon_*$ that would be reached as $\tobs \to\infty$. For the ONC, \citet{kim19a} find that the cluster is virialised and not expanding, which suggests that its star formation efficiency could not be too low. We have no direct dynamical evidence that the same is true for NGC 6530, but given its overall similarity with the ONC, this seems likely to be the case for it as well. The requirement that the star formation efficiency not be ``too low'' is somewhat difficult to quantity: when gas is removed from a protocluster rapidly compared to its dynamical time, loss of more than $\approx 70\%$ of the mass always leads to complete unbinding \citep{kroupa01b}. However, the age distributions in the ONC and NGC 6530 imply that star formation, and presumably mass removal, have been ongoing for significantly longer than a free-fall time, and for sufficiently adiabatic gas removal, stars can remain bound down to arbitrarily small star formation efficiencies. Moreover, in order to match the observation that most stars do not form as part of bound clusters, we require that only a small fraction of the stars remain bound, and thus we do not want the efficiency to be too high. Given our uncertainties, we adopt a relatively mild prior, which disfavours efficiencies below 5\%. Formally, we apply a prior $p_{\rm prior}(\epsilon_{*,\rm obs}) \propto \exp\{-[0.05/\min(\epsilon_{\rm *,obs}, 0.05)]^2\}$. For the purpose of calculating $\epsilon_*$, we adopt $\eta = 1$, corresponding to 50\% instantaneous star formation efficiency, for all models, and a 50\% final star formation efficiency for all but the CBD {and GCD} models. By allowing $\epsilon_{\rm *,clust}$ to be small compared to $\epsilon_*$, we are allowing for the possibility that the clusters are observed early in the formation process, when $M_g\gg M_*$.

Finally, we note that the ONC is also observed to have a small gas fraction at the present day \citep{da-rio14a}, $f_{\rm g, clust} \equiv \Mg(\tobs) / [(\Mg(\tobs) + \Ms(\tobs)] \ll 1$. While in principle this could serve as an additional prior, we lack quantitative constraints on the gas fraction in NGC 6530, and, with the exception of CBD {and GCD}, none of our models contains an explicit treatment of gas clearing. For this reason, we will report $f_{\rm g,clust}$ for our fits, but we will not impose any restrictions on it as a prior.

\begin{table*}
\caption{
\label{tab:fit_params}
Best fit parameters obtained by comparing each model to the observed distribution of stellar ages in the ONC and NGC 6530.  The quantities listed as ``Fit parameters'' are those directly constrained in the fit, while ``Derived parameters'' are calculated from the fit parameters. Values are specified in the form $p(50)_{p(16)-p(50)}^{p(84)-p(50)}$, where $p(q)$ is the $q$th percentile of the marginalised posterior PDF for that quantity. Times expressed as logarithms are in Myr.
}
\begin{tabular}{c@{\qquad\qquad}ccc@{\qquad\qquad}ccc}
\hline\hline
Model & \multicolumn{3}{c}{Fit parameters} & \multicolumn{3}{c}{Derived parameters} \\
      &  $\log\tobs$  &  $\log\tsf$  &   Other  &  $\log\eff$  &   $f_{\rm g,clust}$  & $\log\epsilon_{\rm *,clust}$  \\
      &  [Myr]        &  [Myr]  \\
\hline
\multicolumn{7}{c}{ONC} \\ \hline
ST  &   $0.78_{-0.02}^{+0.02}$  &   $1.72_{-0.14}^{+0.16}$  &   --  &    $-2.24_{-0.16}^{+0.14}$   &   $0.94_{-0.02}^{+0.02}$   &   $-1.26_{-0.15}^{+0.13}$   \\ [1.5ex]
CB, $p=0$  &   $0.92_{-0.02}^{+0.03}$  &   $1.46_{-0.24}^{+0.17}$  &    $\log\ta = 0.89_{-0.04}^{+0.03}$   &    $-1.98_{-0.17}^{+0.24}$   &   $0.92_{-0.05}^{+0.02}$   &   $-1.15_{-0.15}^{+0.21}$   \\ [1.5ex]
CB, $p=3$  &   $1.13_{-0.04}^{+0.04}$  &   $-0.28_{-0.19}^{+1.33}$  &    $\log\ta = 1.12_{-0.04}^{+0.04}$   &    $-0.24_{-1.33}^{+0.19}$   &   $0.12_{-0.07}^{+0.73}$   &   $-0.33_{-0.55}^{+0.02}$   \\ [1.5ex]
CBD, $p=3$  &   $1.12_{-0.07}^{+0.04}$  &   $0.11_{-0.47}^{+1.35}$  &    $\log\ta = 1.11_{-0.17}^{+0.04}$, $\log\phid = 1.48_{-1.10}^{+0.38}$   &    $-0.63_{-1.35}^{+0.47}$   &   $0.02_{-0.02}^{+0.89}$   &   $-0.44_{-0.67}^{+0.09}$   \\ [1.5ex]
GC  &   $0.94_{-0.04}^{+0.04}$  &   $1.34_{-0.11}^{+0.78}$  &    $\log\tc = 0.94_{-0.05}^{+0.07}$, $\xi = 1.00_{-0.06}^{+0.18}$   &    $-1.00_{-0.72}^{+0.07}$   &   $0.00_{+0.00}^{+0.92}$   &   $-0.30_{-0.82}^{+0.00}$   \\ [1.5ex]
\multirow{2}{*}{GCD  }&   \multirow{2}{*}{$0.99_{-0.05}^{+0.05}$  }&   \multirow{2}{*}{$1.92_{-0.49}^{+0.43}$  }&    $\log\tc = 0.99_{-0.05}^{+0.05}$, $\log\tfb = 0.81_{-0.64}^{+0.15}$, &    \multirow{2}{*}{$-1.33_{-0.46}^{+0.31}$   }&   \multirow{2}{*}{$0.00_{+0.00}^{+0.64}$   }&   \multirow{2}{*}{$-0.74_{-0.42}^{+0.35}$   }\\ [0.5ex]& & & $\log\phid = 0.60_{-0.44}^{+0.53}$, $\xi = 1.00_{+0.00}^{+2.89}$   \\ [1.5ex]
IE  &   $1.08_{-0.04}^{+0.04}$  &   $4.76_{-0.61}^{+0.47}$  &    $\delta = 2.68_{-0.42}^{+0.23}$   &    $-1.31_{-0.08}^{+0.09}$   &   $0.92_{-0.06}^{+0.03}$   &   $-1.13_{-0.16}^{+0.23}$   \\ [1.5ex]
\hline
\multicolumn{7}{c}{NGC6530} \\ \hline
ST  &   $0.52_{-0.02}^{+0.02}$  &   $1.64_{-0.16}^{+0.13}$  &   --  &    $-2.25_{-0.13}^{+0.16}$   &   $0.96_{-0.02}^{+0.01}$   &   $-1.44_{-0.12}^{+0.15}$   \\ [1.5ex]
CB, $p=0$  &   $0.66_{-0.02}^{+0.02}$  &   $1.35_{-0.12}^{+0.18}$  &    $\log\ta = 0.64_{-0.03}^{+0.03}$   &    $-1.95_{-0.18}^{+0.12}$   &   $0.95_{-0.02}^{+0.02}$   &   $-1.29_{-0.16}^{+0.11}$   \\ [1.5ex]
CB, $p=3$  &   $0.92_{-0.03}^{+0.02}$  &   $1.03_{-0.39}^{+0.23}$  &    $\log\ta = 0.90_{-0.03}^{+0.03}$   &    $-1.63_{-0.23}^{+0.39}$   &   $0.91_{-0.11}^{+0.03}$   &   $-1.10_{-0.18}^{+0.32}$   \\ [1.5ex]
CBD, $p=3$  &   $0.88_{-0.27}^{+0.03}$  &   $0.52_{-0.36}^{+1.04}$  &    $\log\ta = 0.87_{-0.97}^{+0.03}$, $\log\phid = 1.84_{-1.59}^{+0.12}$   &    $-1.13_{-1.04}^{+0.36}$   &   $0.43_{-0.34}^{+0.53}$   &   $-0.77_{-0.61}^{+0.20}$   \\ [1.5ex]
GC  &   $0.83_{-0.05}^{+0.04}$  &   $2.23_{-1.03}^{+0.20}$  &    $\log\tc = 0.86_{-0.09}^{+0.04}$, $\xi = 2.64_{-1.64}^{+0.67}$   &    $-1.50_{-0.21}^{+0.47}$   &   $0.93_{-0.93}^{+0.02}$   &   $-1.21_{-0.17}^{+0.91}$   \\ [1.5ex]
\multirow{2}{*}{GCD  }&   \multirow{2}{*}{$0.87_{-0.04}^{+0.04}$  }&   \multirow{2}{*}{$2.40_{-0.17}^{+0.18}$  }&    $\log\tc = 0.88_{-0.04}^{+0.04}$, $\log\tfb = 0.58_{-0.67}^{+0.19}$, &    \multirow{2}{*}{$-1.28_{-0.82}^{+0.19}$   }&   \multirow{2}{*}{$0.89_{-0.89}^{+0.04}$   }&   \multirow{2}{*}{$-1.29_{-0.17}^{+0.14}$   }\\ [0.5ex]& & & $\log\phid = 0.87_{-0.22}^{+0.28}$, $\xi = 8.20_{-7.20}^{+1.34}$   \\ [1.5ex]
IE  &   $0.85_{-0.02}^{+0.02}$  &   $4.60_{-0.20}^{+0.20}$  &    $\delta = 2.95_{-0.08}^{+0.04}$   &    $-1.14_{-0.05}^{+0.04}$   &   $0.95_{-0.02}^{+0.02}$   &   $-1.29_{-0.17}^{+0.11}$   \\ [1.5ex]
\hline\hline
\end{tabular}
\begin{tablenotes}
\item $^{(a)}$ We derive $\eff$ as follows: for models ST, CB, and CBD, we use \autoref{eq:tsf} with $\tff$ set equal to the observed value in NGC 6530 or the ONC. For model IE, we report the time-averaged value $\overline{\epsilon}_{\rm ff}$ given by \autoref{eq:eff_mean}. Finally, for model{s GC and GCD} we use the procedure described in \autoref{note:ghc_xi}. In all cases our numerical value is for $\eta = 1$, and $\eff$ obeys the scaling $\eff \propto 1/(1+\eta)$.
\end{tablenotes}
\end{table*}

\begin{figure*}
\includegraphics[width=\textwidth]{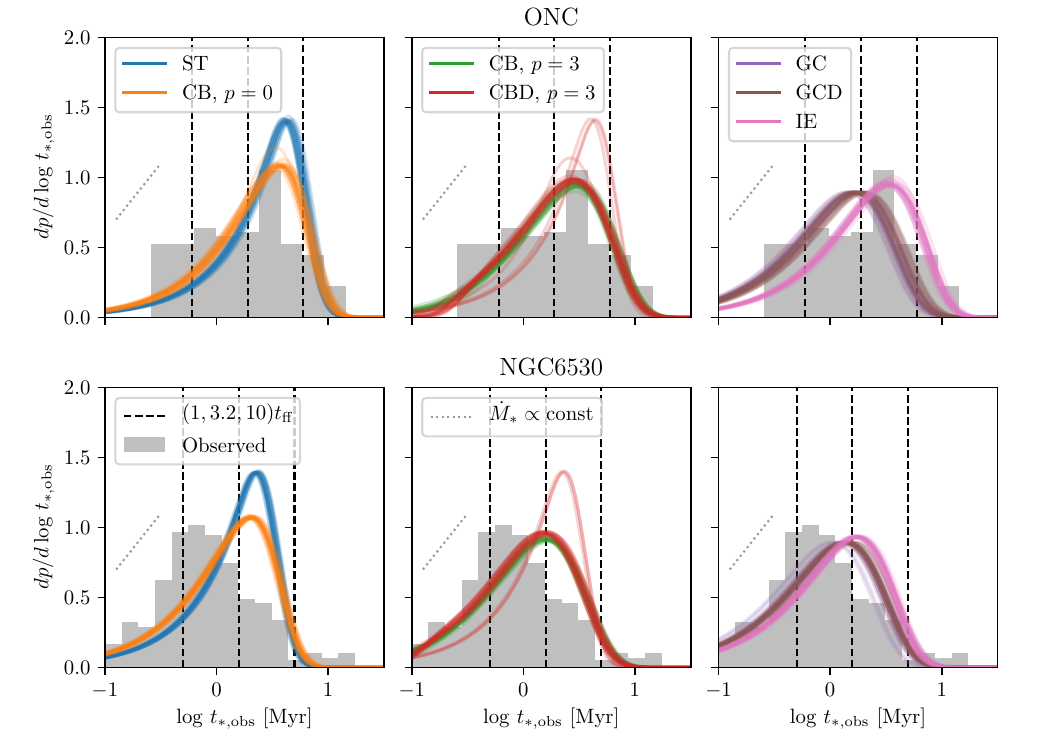}
\caption{
\label{fig:age_fits}
Distribution of observed stellar ages $dp/d\log t_{\rm *,obs}$ in the ONC (top) and NGC 6530 (bottom). In all panels, the coloured lines represent 20 random samples from the final iteration of the MCMC for each of the models, as indicated in the legend. Grey histograms show the observed distribution, and are the same in every panel. The dashed vertical lines indicate $1$, $3.2$, and $10\times{}$ the observed free-fall time, as indicated. {The dotted grey lines, which we provide to guide the eye, are lines of slope unity, which, given the logarithmic age bins, corresponds to a constant star formation rate.}
}
\end{figure*}

\subsubsection{Results}

Having defined the likelihood function and priors, we use the package \textsc{emcee} \citep{foreman-mackey13a} to perform a Markov Chain Monte Carlo (MCMC) calculation to determine the posterior probability distribution for all the free parameters in each model as compared to the data; for the CB model we consider both a case with our fiducial value, $p=3$, and one with $p=0$, as predicted for late stages of collapse by \citet{lee16b,lee16c}. For this calculation we use 100 walkers and perform 1000 MCMC steps; visual inspection of the chains indicates that this is more than adequate for convergence. We report the marginalised posterior PDFs, which we derive from the final 800 steps (i.e., we use 200 steps as a burn in period), in \autoref{tab:fit_params}, and show the fits in \autoref{fig:age_fits}. We provide full posterior PDF distributions of all variables as Supplementary material (online). In \autoref{tab:fit_params} we also report three additional. derived quantities for each model, which are helpful in interpreting the results: the star formation rate per free-fall time $\eff$, the present-day gas fraction $f_{\rm g,clust}$, and the present-day star formation efficiency $\epsilon_{*,\rm clust}$.

Our analysis allows a few immediate conclusions. First, examining \autoref{fig:age_fits}, it is clear that the ST and CB ($p=0$) models provide a poor description of the data in both NGC 6530 and the ONC. The underlying reason is that ST always produces a star formation rate that is highest at the start of star formation and then tapers; CB with $p=0$ has a star formation rate that accelerates with time only weakly. Both models therefore predict a stellar age distribution that is peaked toward the oldest ages, contrary to what we observe. The MCMC attempts to compensate for this effect by favouring large star formation timescales $\tsf$, so that as little gas is converted to stars as possible and the star formation rate falls off due to gas depletion as little as possible; this is also why both models have very high present-day gas fraction $f_{\rm g,clust}$ and very low present-day star formation efficiency $\epsilon_{\rm *,clust}$.

The IE model provides a better fit to the data, but in order to do so the fit is driven to values of $\delta$, the acceleration parameter, far from the theoretically-preferred value $\delta = 1$. Indeed, the only reason $\delta$ does not go even higher is that our priors do not allow $\delta > 3$. Physically, this is because the model has difficulty producing a star formation history that extends for many free-fall times but also accelerates strongly at late times, unless $\delta$ is very large. The existence of a reasonably population of stars with ages approaching $\sim 10\tff$ requires that the $\tobs$ not be too small, but then if $\delta$ is close to unity, too much gas is consumed at early times to allow the star formation rate to accelerate at later times. Thus in order to fit the data, the model requires a much larger value of $\delta$, which more strongly suppresses star formation at early times.

The most successful models are CB, CBD, GC{, and GCD. Though none of the models are able to reproduce the full age distribution in great detail, all four} produce accelerating star formation that is in reasonable agreement with the observed age distribution, with an accretion time (for CB or CBD) or a collapse time (for GC {and GCD}) that is nearly equal to the age of the oldest stars present, and to our best estimate for the age of the system as a whole. {Given that our model for the uncertainties in stellar age estimates is almost certainly too simplistic, this is probably the best level of agreement for which it is reasonable to hope.} The posterior distribution of dimensionless star formation efficiency $\eff$ in these models is extremely broad, mainly because the star formation history is relative insensitive to gas consumption, and instead reflects the accumulation of additional mass (at a rate in good agreement with that predicted by \citealt{goldbaum11a}) in CB or CBD, or to the overall increase in the density and thus decrease in the free-fall time in GC {or GCD}. Interestingly, in the ONC all three models either admit or require that the present-day gas fraction $f_{\rm g,clust}$ be small, consistent with the observations of \citet{da-rio14a}, though we did not explicitly impose this as a prior.

\subsection{$\eff$ from YSO counts}
\label{ssec:atlasgal}

\subsubsection{Data set}

The next observational test to which we subject our models is the relationship between gas and YSOs in the gaseous objects that are the likely progenitors of star clusters. As discussed in the \autoref{sec:intro}, estimates of $\eff$ based on YSO counts cluster around $\approx 0.01$ in all observed star-forming regions, with small scatter. While this would seem to straightforwardly and directly constrain $\eff$, a number of authors have suggested that this is not the case due to biases introduced by the methodologies of the measurement. For example, \citet{lee16a} argue that some measurements preferentially select clouds early in their evolution, when, according to \citeauthor{lee16a}'s favoured IE model, $\eff$ is smaller than its time-averaged value. Similarly, \citet{vazquez-semadeni19a} favour a {GCD} model and argue that estimates of $\eff$ may be erroneous in clouds because a count of the number of YSOs present implicitly integrates the star formation rate over some period of time into the past, when the free-fall time was longer than the value we measure at the present day. We are in a position to test both these hypotheses, by directly modelling the observed distribution of $\eff$ values produced by our cluster formation models.

We take our measured distribution of $\eff$ values from \citet{heyer16a}, who identify class 0/I YSOs within and measure $\eff$ for gas clumps identified in the ATLASGAL survey \citep{schuller09a, csengeri14a}. {We use \citeauthor{heyer16a}'s IMF-corrected estimates of $\eff$, which account statistically for the fact that their YSO catalogs begin to suffer from incompleteness for protostars smaller than $\approx 2$ $M_\odot$.} This is the largest ($N=517$)\footnote{For some of this sample \citet{heyer16a} do not detect any YSOs, and thus only obtain an upper limit on $\eff$. For the purposes of our analysis we take the value of $\eff$ in these clumps to be equal to the stated $2\sigma$ upper limit.} and most complete sample of $\eff$ measurements in the literature, and the ATLASGAL clumps that it targets are very similar to the ONC and NGC 6530 in terms of mass, density, and free-fall time, making the data well-suited to the task of using both the cluster star formation history and the $\eff$ distribution together, as we do below. Specifically, the mean free-fall time of the ATLASGAL clumps is $0.3$ Myr, very similar to the observed free-fall times of $0.5$ Myr and $0.6$ Myr in NGC 6530 and the ONC. Thus the ATLASGAL sample very likely represents a survey of YSOs in objects that are will become clusters like NGC 6530 or the ONC, just at a slightly earlier evolutionary phase. However, we do note that the distribution of $\eff$ values obtained by \citet{heyer16a} is qualitatively quite similar to those obtained from other samples that also use YSO counts for objects at a range of size and density scales \citep[e.g.,][]{evans14a, ochsendorf17a}.

\subsubsection{Likelihood function}

As in \autoref{ssec:sf_hist}, to compare to the models to the observations we require a likelihood function that gives the probability density of the data given the model, which must properly account for averaging of the star formation rate over a finite time interval, potential biases in the sample, and observational errors. First consider the issue of averaging over a finite time. The \citet{heyer16a} data set on which we focus estimates the star formation rate (SFR) based on number counts of class 0/I YSOs, a phase that lasts for a time $t_{\rm YSO} \approx 0.5$ Myr \citep{evans09a, gutermuth09a}. We can therefore define an appropriately time-averaged $\eff$ for our models as
\begin{equation}
\label{eq:eff_avg}
\epsilon_{\mathrm{ff,avg}}(t, \Delta t) = \frac{\left[M_*(t) - M_*(t - \Delta t)\right]/\Delta t}{\Mg(t)/\tff(t)},
\end{equation}
where $t$ is the time of observation, and $\Delta t = 0.5$ Myr is the window over which the SFR is averaged.

As with our treatment of stellar ages, we must consider not only biases (in this case introduced by averaging over a finite time), but observational errors. Errors in $\eff$ measurements are significantly more poorly modelled than errors in stellar age distributions, and involve subtleties such as making an IMF-based correction for the presence of protostars too dim to be detected. Given our ignorance, we will adopt a simple lognormal functional form, i.e., in a cloud with a true (time-averaged) logarithmic star formation efficiency $\log\epsilon_{\mathrm{ff,avg}}$, the distribution of observationally-inferred values $\log\effobs$ will be distributed as a Gaussian of width $\sigma_{\log \eff}$ centred on $\log\epsilon_{\mathrm{ff,avg}}$. That is, given a true (time-averaged) efficiency per free-fall time $\epsilon_{\mathrm{ff,avg}}$, the distribution of observationally-estimated star formation efficiency per free-fall time is
\begin{eqnarray}
\lefteqn{
f(\log\effobs \mid \log \epsilon_{\mathrm{ff,avg}}) =
}
\nonumber \\
& &
 \frac{1}{\sqrt{2\pi}\sigma_{\log\eff}}
\exp\left[-\frac{\left(\log\epsilon_{\mathrm{ff,avg}} - \log\effobs\right)^2}{2\sigma_{\log\eff}^2}\right].
\end{eqnarray}
The value of the dispersion $\sigma_{\log\eff}$ is not well known, but we will see below that it is not necessary to adopt a model for $\sigma_{\log\eff}$; instead we can leave $\sigma_{\log\eff}$ as a parameter to be fit along with other model parameters.

\begin{figure*}
\includegraphics[width=\textwidth]{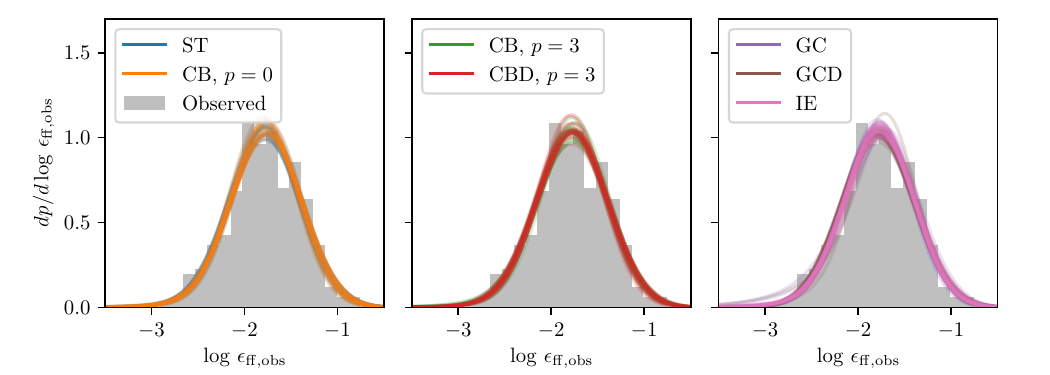}
\caption{
\label{fig:eff_dist}
Distribution of observed star formation efficiencies $\log \epsilon_{\rm ff,obs}$. Grey histograms show the distribution observed by \citet{heyer16a} for the ATLASGAL sample, and are the same in every panel. Coloured lines represent 20 random samples from the final iteration of the MCMC fit for each model, as indicated in the legend.
}
\end{figure*}

Now consider a cloud observed at some time $t$ during its evolution, with an instantaneous time-averaged star formation efficiency $\epsilon_{\mathrm{ff,avg}}(t, \Delta t)$. The distribution of observed efficiencies for this cloud is $f(\log\effobs \mid \log \epsilon_{\mathrm{ff,avg}}(t,\Delta t))$. If we have a population of such clouds, each observed at random times $t$ between the onset of star formation at $t=0$ and some maximum time $t_{\rm max}$, then the distribution of observed $\effobs$ values for the population is simply the average of $f(\log\effobs \mid \log \epsilon_{\mathrm{ff,avg}}(t,\Delta t))$ over all possible times $t$ at which the clouds could be observed, i.e.,
\begin{equation}
\frac{dp}{d\log\effobs} = \frac{1}{t_{\rm max}} \int_0^{t_{\rm max}} f(\log\effobs \mid \log \epsilon_{\mathrm{ff,avg}}(t,\Delta t)) \, dt.
\end{equation}
The choice of maximum time $t_{\rm max}$ is somewhat subtle. In simple models where $\Mg$ reaches 0 in finite time, such as the GC model, one can simply take $t_{\rm max}$ to be the time for which $\Mg(t_{\rm max}) = 0$. However, we are interested in comparing to a more general class of models where $\Mg$ may not go to exactly 0 at finite time. To choose a reasonable $t_{\rm max}$, we note that studies of $\eff$ based on YSO counts always select YSOs and gas clouds within the same area on the sky, which limits the phase of evolution to which they are sensitive: as clusters evolve and begin to clear their gas, stars inevitably cease to be surrounded by molecular gas, so clouds that have cleared most of their gas are not included in YSO counting surveys. Our simple zero-dimensional models cannot capture this effect directly, but we crudely mimic it by choosing our time interval to correspond to that over which $\Mg/M_* > 1$, i.e., when the stellar mass has not yet exceeded the gas mass. We therefore take $t_{\rm max}$ to be defined implicitly by the condition $\Mg(t_{\rm max})/M_*(t_{\rm max}) = 1$. We have verified that varying the value of $\Mg/M_*$ we use to define our time interval by a factor of ten in either direction not change the results substantially.

Given the preceding discussion, we have now write down the log likelihood function for a set of observed $\eff$ values is 
\begin{equation}
\log\mathcal{L} = \sum_{i=1}^N \log \left(\frac{dp}{d\log\effobs}\right)_{\effobs = \epsilon_{{\rm ff},i}},
\end{equation}
where $\epsilon_{{\rm ff},i}$ is the $i$th observed value of $\eff$, and there are $N$ measurements in total. We use this likelihood function with \textsc{emcee} to obtain posterior PDFs for the parameters for the same models as in \autoref{ssec:sf_hist}. As in our analysis of the stellar age distribution, we use priors that are flat in the logarithm of positive-definite quantities, and flat in value for other quantities; the allowed parameter range is identical to that used in \autoref{ssec:sf_hist}. In addition to the parameters included there, we must also fit for $\eta$, $\sigma_{\log\eff}$ and $\xi$ (for model GC {and GCD}), since, while these do not affect the distribution of stellar ages, they do affect the distribution of observed $\eff$ values. For $\eta$ our prior is flat in log from $0.01 - 10$, and for $\sigma_{\log\eff}$ it is flat in log from $0.01 - 10$. We must also choose a value for the free-fall time, since this sets the ratio $\Delta t/\tff$, which determines how much the observed $\eff$ distribution is biased by averaging the star formation rate over a finite time. As noted above, the mean value of $\tff$ in the ATLASGAL sample is $0.3$ Myr, and the dispersion around this is small ($0.26$ dex), so we use $\tff = 0.3$ Myr for our analysis of all models except GC {and GCD; these} models sweep through all values of $\tff$ from $\tffz$ to 0, so for this case we impose as a prior the requirement that $\tffz > 0.3$ Myr, i.e., the collapse must start from a state that is no denser than the observed ATLASGAL clumps.

\subsubsection{Results}

\begin{table*}
\caption{
\label{tab:fit_eff}
Best fit parameters obtained by comparing each model to the observed distribution of measured $\eff$ values in ATLASGAL clumps \citep{heyer16a}.
}
\begin{tabular}{c@{\qquad\qquad}cccc@{\qquad\qquad}c}
\hline\hline
Model & \multicolumn{4}{c}{Fit parameters} & Derived parameters \\
     &   $\log\sigma_{{\log\eff}}$  &   $\log\eta$$^{{(a)}}$  &  $\log\tsf$  &   Other  &  $\log\eff$$^{{(b)}}$  \\
     &   [dex]                        &                 &  [Myr] \\ \hline
ST  &   $-0.79_{-0.01}^{+0.01}$  &   $-0.72_{-0.86}^{+0.86}$  &   $1.18_{-0.30}^{+0.06}$  &   --  &    $-1.78_{-0.02}^{+0.02}$   \\ [1.5ex]
CB, $p=0$  &   $-0.79_{-0.02}^{+0.02}$  &   $-0.53_{-0.93}^{+0.82}$  &   $1.13_{-0.36}^{+0.10}$  &    $\log\ta = 1.45_{-0.81}^{+0.38}$   &    $-1.76_{-0.02}^{+0.02}$   \\ [1.5ex]
CB, $p=3$  &   $-0.79_{-0.02}^{+0.02}$  &   $-0.65_{-0.89}^{+1.02}$  &   $1.13_{-0.44}^{+0.08}$  &    $\log\ta = 1.77_{-0.28}^{+0.16}$   &    $-1.74_{-0.02}^{+0.02}$   \\ [1.5ex]
CBD, $p=3$  &   $-0.79_{-0.02}^{+0.02}$  &   $-0.66_{-0.88}^{+0.92}$  &   $1.13_{-0.38}^{+0.08}$  &    $\log\ta = 1.78_{-0.26}^{+0.15}$, $\log\phid = 1.02_{-0.70}^{+0.68}$   &    $-1.74_{-0.02}^{+0.03}$   \\ [1.5ex]
GC  &   $-0.78_{-0.02}^{+0.02}$  &   $-0.08_{-0.82}^{+0.70}$  &   $2.53_{-0.86}^{+0.89}$  &    $\log\tc = 1.58_{-0.50}^{+0.31}$, $\log\xi = -0.00_{-0.66}^{+0.72}$   &    $-1.77_{-0.02}^{+0.03}$   \\ [1.5ex]
\multirow{2}{*}{GCD  }&   \multirow{2}{*}{$-0.79_{-0.03}^{+0.02}$  }&   \multirow{2}{*}{$0.17_{-0.65}^{+0.54}$  }&   \multirow{2}{*}{$2.47_{-0.88}^{+0.72}$  }&    $\log\tc = 1.39_{-0.57}^{+0.43}$, $\log\xi = 0.22_{-0.72}^{+0.52}$   &    \multirow{2}{*}{$-1.77_{-0.03}^{+0.06}$   } \\ [0.5ex]   & & & & $\log\tfb = -0.45_{-1.04}^{+1.02}$, $\log\phid = 0.82_{-0.57}^{+0.67}$   \\ [1.5ex]
IE  &   $-0.79_{-0.02}^{+0.02}$  &   $-0.69_{-0.88}^{+0.79}$  &   $1.25_{-0.27}^{+0.12}$  &    $\delta = 0.06_{-0.04}^{+0.09}$   &    $-1.76_{-0.03}^{+0.03}$   \\ [1.5ex]
\hline\hline
\end{tabular}
\begin{tablenotes}
\item Formatting is identical to that used in \autoref{tab:fit_params}.
\item $^{(a)}$ The median and percentile values we report for $\log\eta$ are strongly affected by our prior $\log\eta > -2$. All models with $\eta \ll 1$ are essentially identical, so our analysis cannot distinguish them; thus the values we report should be read as providing an upper limit at the reported 84th percentile, rather than a meaningful central estimate.
\item $^{(b)}$ The value of $\eff$ we report here is the true value defined by the instantaneous star formation rate, not the time-averaged value $\epsilon_{\mathrm{ff,avg}}$ defined by \autoref{eq:eff_avg}. For model IE, we report the time-averaged value $\overline{\epsilon}_{\rm ff}$ given by \autoref{eq:eff_mean}. We compute $\eff$ as described in the notes to \autoref{tab:fit_params}.
\end{tablenotes}
\end{table*}

We show models evaluated using samples drawn from the MCMC chains in \autoref{fig:eff_dist}, and report the posterior PDFs of all parameters in \autoref{tab:fit_eff}. The results show that all the models we consider can fit the observed $\eff$ distribution quite well, but that both $\eff$ and the level of observational error are \textit{very} tightly constrained by the observations; $\eff$ is required to be of order a few percent, and $\sigma_{\log\eff}$ to be approximately 0.15 dex. Indeed, the models even constrain $\eta$ not be too large, since otherwise rapid mass removal means that the gas mass is able to change significantly over the time-averaging interval $\Delta t$, which in turn would broaden the observed $\eff$ distribution more than the data allow. Thus, despite the hypothesis in the literature that measured $\eff$ distributions are biased because they average over a finite time interval and thus miss changes in the free-fall time \citep[e.g.,][]{vazquez-semadeni19a}, or that they miss periods of efficient star formation \citep[e.g.,][]{lee16a}, we do not obtain significantly looser constraints on the value of $\eff$ when we explicitly put those possibilities into our model. 


\begin{figure}
\includegraphics[width=\columnwidth]{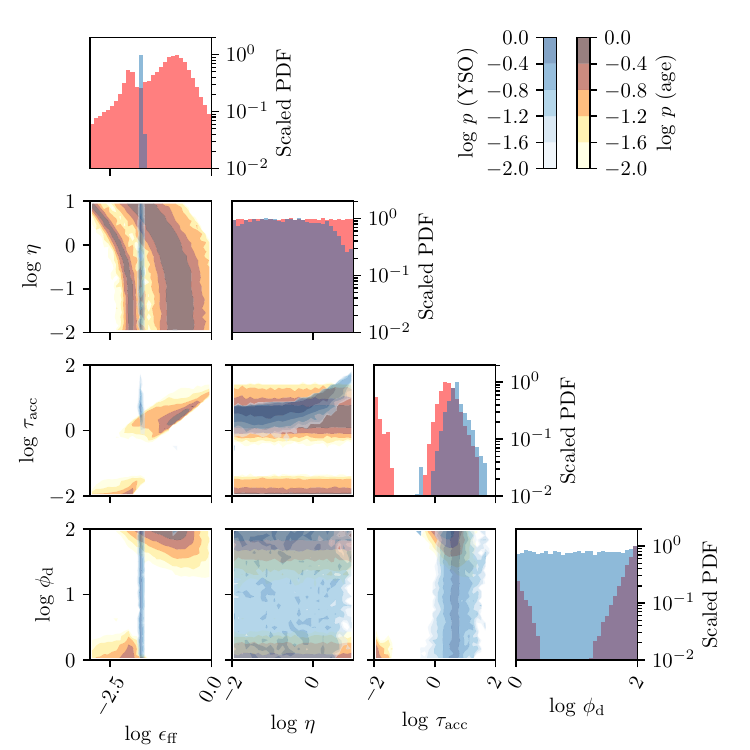}
\caption{
\label{fig:combined_cbd}
Corner plot showing the posterior PDF for the dimensionless parameters of the CBD model ($\eff$, $\eta$, $\tau_{\rm acc}$, and $\phid$), derived using the distribution of stellar ages in NGC 6530 (red colours) and the counts of YSOs in ATALASGAL clumps (blue colours). In the panels on the bottom left corner, contours show 2D marginal posterior PDFs for each combination of variables, as indicated on the axes. Histograms along the central diagonal show 1D marginal posterior PDFs for each variable. PDFs in all panels are scaled so that the maximum is unity.
}
\end{figure} 

\begin{figure}
\includegraphics[width=\columnwidth]{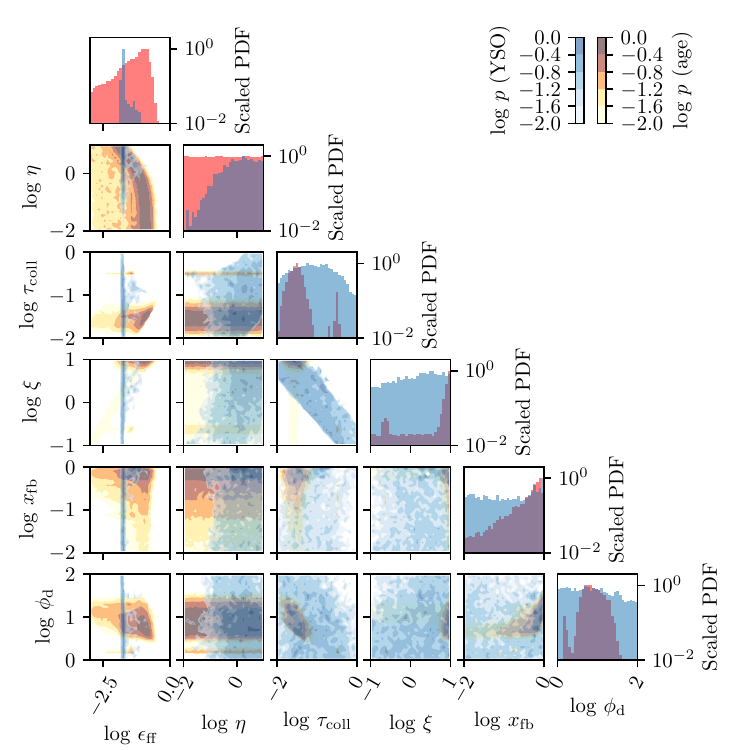}
\caption{
\label{fig:combined_gcd}
{Same as \autoref{fig:combined_cbd}, but showing the GCD model and its dimensionless parameters. Note that we show $x_{\rm fb} = \tfb/\tc$ rather than $\taufb = \tfb/\tsf$, because the former quantity is more helpful for the discussion that follows.}
}
\end{figure}

\begin{figure}
\includegraphics[width=\columnwidth]{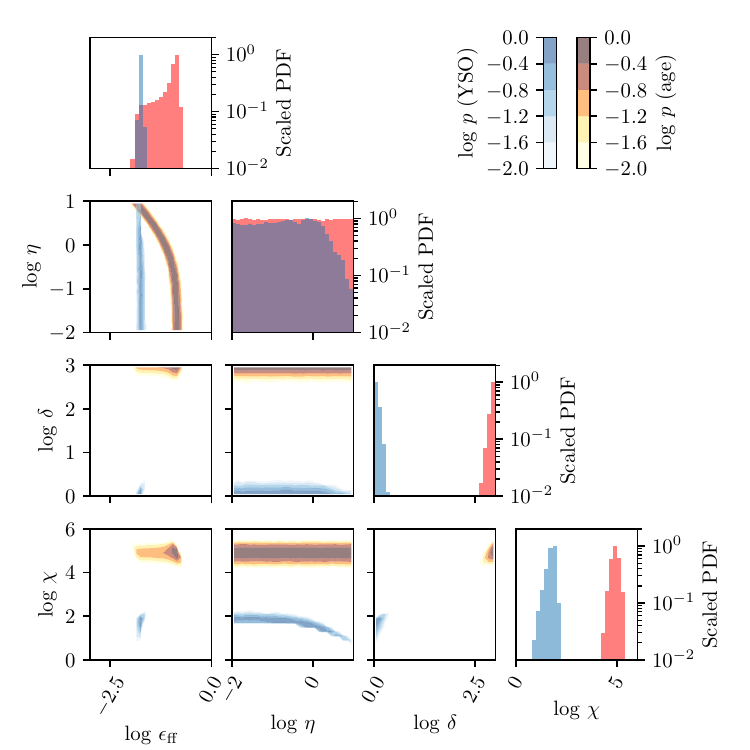}
\caption{
\label{fig:combined_ie}
Same as \autoref{fig:combined_cbd}, but showing the IE model and its dimensionless parameters.
}
\end{figure}

\begin{figure}
\includegraphics[width=\columnwidth]{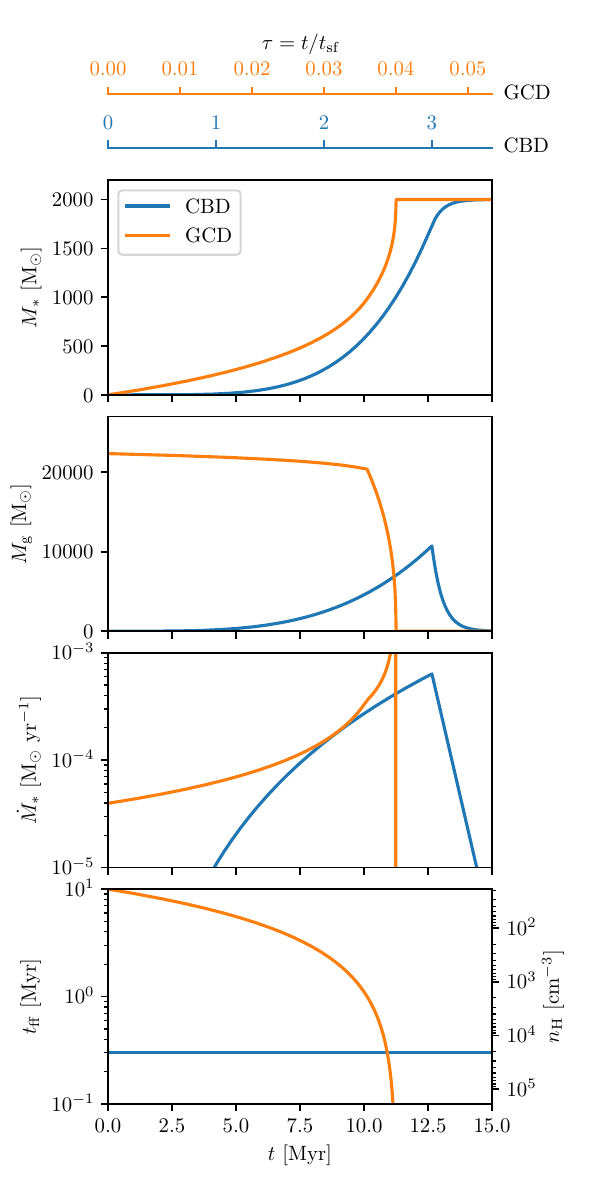}
\caption{
\label{fig:best_fit}
{
Example histories of stellar mass, gas mass, star formation rate, and free-fall time / density for the two best-fitting models, CBD and GCD, scaled to mass and time scales typical of the ATLASGAL sample; the exact parameters used to construct these models are described in \autoref{ssec:combined}. The right axis in the bottom panel shows number density of H nuclei, computed assuming a mean mass of $1.4m_{\rm H}$ per H nucleon. The bottom horizontal axis shows physical time in Myr, while the top two axes show dimensionless time $\tau = t/\tsf$; this is different for the CBD and GCD models because the star formation timescale $\tsf$ is different in the two models. 
}
}
\end{figure}

\subsection{Combined constraints}
\label{ssec:combined}

Having examined the constraints we can deduce from the distribution of stellar ages and the YSO-gas correlation individually, we now ask whether these constraints are compatible. That is, do there exist a set of parameters for a given model such it can simultaneously reproduce the observed stellar age distribution in young clusters and the YSO count in protoclusters? To answer this question, we use our MCMC samples to compute the dimensionless parameters -- $\eff$, $\eta$, etc. -- that characterise each proposed model, using the constraints from both the stellar age distribution and YSO counts. {We focus only on the dimensionless parameters, since, while the star clusters for which we have examined the stellar age distribution and the ATLASGAL clumps are similar in terms of mass and free-fall time, they are not completely identical, and thus we do not expect the dimensional parameters (e.g., free-fall time or collapse time) to match exactly.} We plot the posterior PDFs of the dimensionless parameters for models CBD, {GCD,} and IE in \autoref{fig:combined_cbd}, \autoref{fig:combined_gcd}, and \autoref{fig:combined_ie}, respectively.\footnote{For parameters that cannot be constrained by the stellar age distribution, we take the posterior PDF derived from stellar ages to be equal to the flat prior we use for these variables when analysing the ATLASGAL data.} These plots use the posterior PDFs derived from the stellar age distribution in NGC 6530, since it is a somewhat larger data set, but the results for the ONC are qualitatively similar. We omit ST and CB ($p=0$) from this comparison because we have already determined that these models provide poor fits to the stellar age distribution alone, and we omit CB ($p=3$) {and GC} because {they are} qualitatively similar to CBD {and GCD, respectively,} on the parameters they share. However, the corresponding plots for these clusters are provided in the Supplementary material (online).

Turning first to \autoref{fig:combined_ie}, we immediately see that the IE model has a major difficulty: as discussed in \autoref{ssec:sf_hist} and shown in \autoref{fig:combined_ie}, the stellar ages distributions in NGC 6530 and the ONC are best fit in the context of this model by a star formation efficiency that increases as roughly $\eff \propto t^3$ (or faster, since $\delta = 3$ is the largest allowed by our priors). This is completely at odds with the constraint provided by the ATLASGAL clumps, whose tight relationship between YSOs and gas properties requires that $\eff$ be nearly constant, and thus that $\delta \approx 0$. The physical explanation for this tension is simple: star formation is observed to accelerate based on stellar age distributions, and the IE model interprets this acceleration as a systematic increase in star formation efficiency with time. However, when one observes the gas clumps that are in the process of forming clusters, one finds that the number of YSOs per unit gas mass, normalised by the free-fall time, is nearly constant, completely inconsistent with large variations in star formation efficiency. There is no way to reconcile these two constraints in the context of the IE model, or indeed in any model that assumes the acceleration of star formation is due to an increase in star formation efficiency with time. Instead, the acceleration of star formation must be due either to an increase in the star-forming mass with time (as in CB or CBD) or a decrease in the free-fall time (as in GC {or GCD}). We may therefore rule out the IE model.

The CBD and {CGD} models illustrated respectively in \autoref{fig:combined_cbd} and \autoref{fig:combined_gcd}, on the other hand, show no contradiction between the parameter values demanded by the stellar age distributions and the ATLASGAL clumps. In both sets of models the ATLASGAL data very tightly constrain $\eff$, while setting little constraint on {any other parameters}. Conversely, the stellar age distribution tightly constrains $\tau_{\rm acc}$, $\tau_{\rm coll}$, $\xi$, {$x_{\rm fb}$, and $\phid$}, but provides little restriction on $\eff$. As a result, there is a reasonable parameter space of overlap. 

Thus we find that the joint set of data favour one of two scenarios. {We plot the history of gas and stellar mass, star formation rate, and mean density and free-fall time derived for these two scenarios in \autoref{fig:best_fit}.} In the first, gas accretes as roughly $\dot{M}\propto t^3$ (consistent with the theoretical models of \citealt{goldbaum11a}) and forms stars inefficiently ($\eff \approx 0.01$). Accretion continues for $\sim 1-10$ star formation timescales ($\taua \sim 1-10$), and once it ends, mass is rapidly dispersed by feedback ($\phid \gg 1$). {The precise parameters used for the CBD model shown in \autoref{fig:best_fit} are $\log\eff = -1.75$, $\tau_{\rm acc} = 3$, $\eta = 3$, $\phi_d = 10$; all of these parameters are within the 16th to 84th percentile range allowed by both sets of constraints. The gas and stellar masses in the model, physical time, and star formation rate, can be rescaled arbitrarily by changing the total cloud mass and density, while leaving all the dimensionless parameters (which determine the shape of the curves) fixed. We have scaled the curves shown to values typical of NGC 6530 and the ONC, and of the ATLASGAL clumps: a final stellar mass of 2000 $M_\odot$, and a free-fall time of $0.3$ Myr. The corresponding physical star formation and accretion timescales are $\tsf = 4.2$ Myr and $\ta=12.7$ Myr, respectively.}

In the second scenario, an initially low-density cloud undergoes a global collapse that is fairly rapid compared to the instantaneous free-fall time ($\xi \gtrsim 1$), as might be expected for example in a colliding flow where the collapse is due to external pressure plus gravity rather than gravity alone, but during this collapse it forms stars quite inefficiently ($\eff \approx 0.01$). As a result, the total collapse time is quite small compared to the star formation timescale ($\tauc \lesssim 0.1$), so that most stars form only during the final plunge when the density is running way to infinity -- a value $\tauc < 1$ is required to yield an accelerating star formation history. {The plot shown in \autoref{fig:best_fit} uses $\log\eff = -1.75$, $\tau_{\rm coll} = 0.04$, $\tau_{\rm fb} = 0.036$, $\eta=1.0$ (so $\xi=1.8$), and $\phid = 10$, together with an initial free-fall time $\tffz = 10$ Myr, again falling within the 16th - 84th percentile range of our analysis of NGC 6530 and the ONC; the mass has also been scaled to produce a final stellar mass of 2000 $M_\odot$. The corresponding initial star formation and collapse timescales are $\tsf = 281$ Myr and $\tc = 11.2$ Myr, respectively; the collapse timescale corresponds to a starting density $\approx 20$ cm$^{-3}$, and thus typical of the cold neutral medium (CNM). In this model, the ATLASGAL clouds began their lives as clouds of CNM, and their present-day properties would correspond to a physical state near the point where the blue and orange lines cross in the bottom panel of \autoref{fig:best_fit}.}

\subsection{Global SFR}

We now add an additional constraint to our modelling: the star formation rate of the Milky Way as a whole is $\approx 2$ $M_\odot$ yr$^{-1}$ \citep{chomiuk11a}, so the total star formation rate implied by a successful model must not exceed this value. To see what this implies, we again return to the ATLASGAL sample. As noted above, the mean free-fall time of these objects is $\tff = 0.3$ Myr \citep{heyer16a}, and the total mass of ATLASGAL clumps in the Galaxy is $M_{\rm tot} \approx 1.0\times 10^7$ $M_\odot$ \citep{urquhart18a}. 

\subsubsection{{ST, CB, and CBD}}

The rate at which ATLASGAL clumps form stars is straightforward to calculate in the ST, CB, and CBD models:
\begin{equation}
\SFR = \eff \frac{M_{\rm tot}}{\tff} = 0.33 \left(\frac{\eff}{0.01}\right) \left(\frac{M_{\rm tot}}{10^7\,M_\odot}\right) \left(\frac{\tff}{0.3\,\mathrm{Myr}}\right)^{-1}\,M_\odot\,\mathrm{yr}^{-1},
\label{eq:SFR_CB}
\end{equation}
where we have normalised to the mean free-fall time for the ATLASGAL clumps. Thus if $\eff\approx 0.01$ for these models, as suggested by our analysis so far, the total contribution of the ATLASGAL clumps to the total star formation budget of the Milky Way is $\approx 0.3$ $M_\odot$ yr$^{-1}$, which is $\approx 10\%$ of the total. This is consistent with the upper limit stated above, and in fact suggests a nice consistency: the ATLASGAL clumps are much denser than the mean star-forming region or star cluster (for example, compare to Fig.~9 of \citealt{krumholz19a}), and thus the stars that form within them are much more likely to remain part of a bound cluster than the typical star formed in the Galaxy. If we hypothesise that the ATLASGAL clumps correspond roughly to the bound portion of the star formation in the Galaxy, so our estimate implies that $\sim 10\%$ of all stars formed in bound clusters, that is entirely consistent with the observationally-measured fraction of stars formed in bound clusters in typical spiral galaxies \citep[e.g.,][]{ryon14a, adamo15a, johnson16a, chandar17a}. We caution, however, not to put too much weight on this agreement, since we do not in fact know if the density range that is selected by ATLASGAL corresponds well to the conditions that delineate between bound and unbound star formation.

\subsubsection{{GC, GCD, and IE}}

The {remaining} models require a more refined treatment because $\tff$ and $\eff$ can vary. Since these models do not depend on the magnitude of the mass, we can assume that the entire population of ATLASGAL clouds is born with the same mass and then evolves according to one of these models. Let $\caln_M=d\caln/d\Mg$ be the number of clouds per unit mass, and let $\dot\caln$ be the rate at which clouds are born with a mass $\Mgz$. The equation of continuity for the cloud mass distribution is then
\beq
\ppbyp{\caln_M}{t}+\pbyp{\Mg}\left(\caln_M \dMg\right)=\dot\caln\delta(\Mg-\Mgz),
\eeq
so that in a steady state we have
\beq
-\caln_M\dMg =\dot\caln.
\label{eq:sfrtot}
\eeq
The SFR is then
\beq
\SFR=\int_0^\Mgz \dot M_*\caln_M dM =  {\dot\caln \epsilon_*\Mgz}
\eeq
from \autoref{eq:evol_eq}, since there is no accretion in these models  ($\dMa=0$). {Here $\epsilon_*$ is the final star formation efficiency: $1/(1+\eta)$ in GC or IE, and the value given by \autoref{eq:epsstar_gcd} for GCD.}
Now the total mass in clouds is
\beqa
M_{\rm tot}&=&\int_0^\Mgz \Mg \caln_M dM,\\
&=&\int_0^\infty \Mg\left(\frac{\dot\caln}{\dMg}\right)\left(\frac{d\Mg}{dt}\right)\,dt,\\
&=&\dot\caln \tsf\int_0^\infty \Mg\, d\tau.
\label{eq:mtot}
\eeqa
The star formation rate per unit gas mass is then
\beq
\frac{\SFR}{M_{\rm tot}}= {\frac{\epsilon_*}{\tsf}}\left[\int_0^\infty\frac{\Mg}{\Mgz}\,d\tau\right]^{-1}.
\label{eq:intmg}
\eeq
We can check this by noting that for the ST model it gives the result in \autoref{eq:sfr},
\beq
\SFR=\frac{M_{\rm tot}}{(1+\eta)\tsf}=\eff \frac{ M_{\rm tot}}{\tff}.
\eeq

First consider the IE model. Evaluating the integral in \autoref{eq:intmg} with the aid of \autoref{eq:mg_ie} we find
\begin{eqnarray}
\SFR & = & \frac{\chi^{\delta/(1+\delta)}}{(1+\delta)^{1/(1+\delta)}\Gamma\left(1+\frac{1}{1+\delta}\right)}\effz \frac{M_{\rm tot}}{\tff}\\
& = & \frac{1}{\left(1+\delta\right)\Gamma\left(1+\frac{1}{1+\delta}\right)\Gamma\left(1+\frac{\delta}{1+\delta}\right)} \overline{\epsilon}_{\mathrm{ff}} \frac{M_{\rm tot}}{\tff},
\end{eqnarray}
where in the second step we have made use of \autoref{eq:eff_mean} to rewrite the star formation rate in terms of the mass-averaged star formation efficiency $\overline{\epsilon}_{\rm ff}$. We have already noted that the constraints on $\delta$ arising from stellar age distributions are inconsistent with those derived from YSO counts, but the total star formation rates in both cases are similar. Consulting \autoref{tab:fit_params} and \autoref{tab:fit_eff}, we see that YSO counts give $\overline{\epsilon}_{\rm ff} \approx 0.01$ and $\delta \approx 0$, so overall we obtain $\SFR \approx 0.01 M_{\rm tot}/\tff$. Stellar ages give $\overline{\epsilon}_{\rm ff}\approx 0.05$ and $\delta \approx 3$, which again gives $\SFR \approx 0.01 M_{\rm tot}/\tff$. Thus the global star formation rate predicted by our best-fitting values of the IE model are roughly the same as those obtained in the ST, CB, or CBD models, and is consistent with the global star formation budget of the Milky Way.

Next consider the GC {and GCD} model{s}. In this case, evaluation of the integral in \autoref{eq:intmg} gives
\begin{equation}
\label{eq:sfr_gcd}
\SFR = {f_{\rm GCD}} \left[ \left(\frac{1+\tauc}{\tauc}\right) \eff\right] \frac{M_{\rm tot}}{\tffz}
{
= f_{\rm GCD} \left[\frac{\xi}{2}\left(\frac{1+\tauc}{1+\eta}\right)\right] \frac{M_{\rm tot}}{\tffz}
}
\end{equation}
{where}
\begin{equation}
{
f_{\rm GCD} = \frac{\left(1+\tauc\phid\right)\left[1-\frac{\phid-1}{\phid}\left(1-x_{\rm fb}\right)^{\tauc}\right]}
{1+\tauc\phid \left[1-\frac{\phid-1}{\phid}\left(1-x_{\rm fb}\right)^{\tauc+1}\right]}
}
\label{eq:f_gcd}
\end{equation}
{is the factor by which the SFR is lower in a GCD model than in a GC one due to the extra dispersal at late times; this factor is unity for the GC model, and it also approaches unity for $x_{\rm fb} \to 1$ or $\phid \to1$, in which limits the GCD models reduces to the GC one. The term in square brackets in \autoref{eq:sfr_gcd}, which we have written in two equivalent ways in order to illustrate the limiting behaviour for $\tauc \ll 1$ and $\gg 1$, can be thought of as the ``effective'' $\eff$ of the model. If $\tauc \gg 1$, i.e., clouds collapse slowly compared to their star formation timescale, then clearly this term just approaches $\eff$, \autoref{eq:sfr_gcd} approaches \autoref{eq:SFR_CB}, and this model approaches the behaviour of ST, CB, or CBD. If, on the other hand, $\tauc \ll 1$ so that clouds collapse quickly compared to their star formation timescale, then the term in square brackets approaches $(\xi/2)/(1+\eta)$. This is just the product of the collapse time measured in units of the free-fall time, $(\xi/2)$, and the fraction of the mass converted to stars rather then lost to the wind, $1/(1+\eta)$. The instantaneous value of $\eff$ does not matter in this limit, because all the stars form in the final plunge to infinite density.
}

{We cannot apply this result to the Galaxy as a whole, because both the free-fall time $\tffz$ and the total mass $M_{\rm tot}$ at the start of collapse are unknown -- ATLASGAL tells us only the instantaneous mass of clumps whose density is high enough for them to be included in the catalogue, i.e., those for which $\tff \lesssim 0.3$ Myr. However, we can still apply our model just to the ATLASGAL clumps, simply by interpreting the birth rate $\dot\caln$ as the rate at which clouds become dense enough to be visible to ATLASGAL. Since in the GC and GCD models $\tff$ is monotonically decreasing, we can in this case simply adopt $\tffz = 0.3$ Myr and set $M_{\rm gas}$ to the total mass of the ATLASGAL samples, and then use \autoref{eq:sfr_gcd} to compute the contribution to the Galactic SFR provided by those clumps that are dense and massive enough to fall into the ATLASGAL catalog. This provides a lower limit on the total Galactic SFR.}

{
Inserting the observed mass and free-fall time of the ATLASGAL clumps, we therefore find that the GC and GCD models predict that they should yield a star formation rate
}
\begin{eqnarray}
\lefteqn{
\SFR = 11 \,M_\odot\,\mathrm{yr}^{-1}
}
\nonumber \\
& & {f_{\rm GCD}} \left(\frac{\eff}{0.01}\right) \left(\frac{\tauc}{0.03}\right)^{-1} \left(\frac{M_{\rm tot}}{10^7\,M_\odot}\right) \left(\frac{\tff}{0.3\,\mathrm{Myr}^{-1}}\right).
\label{eq:SFR_ATLASGAL}
\end{eqnarray}
where we have normalised to our best fit value of $\tauc$ based on observed stellar age distributions, and our numerical evaluation assumes $\tauc \ll 1$. We can immediately see that there is a serious problem with the star formation budget in the GC model: for the best-fitting parameters arising from stellar age distributions and YSO counts, the observed ATLASGAL clumps should form stars at nearly five times the total star formation rate of the Galaxy as a whole. The problem becomes even more severe if we recall that ATLASGAL clumps are much denser than the mean density of observed star clusters, and thus likely represent only a small subset of the total star formation in the Galaxy{, i.e., $\mbox{SFR} \gg \mbox{SFR}(<\tff)$.}

\begin{figure}
\includegraphics[width=\columnwidth]{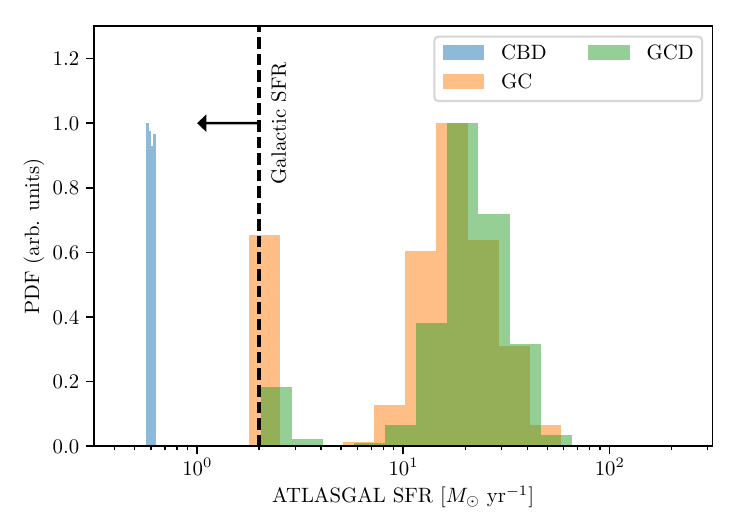}
\caption{
\label{fig:sfr_pdf}
{
Histograms of predicted star formation rates (SFRs) for the gas clumps in the ATLASGAL catalogue, using the CBD, GC, and GCD models with parameters constrained by fitting to the stellar age distribution in young clusters and the number of YSOs per unit gas mass in the ATLASGAL sample. All histograms have been normalised to have a maximum of unity for ease of comparison. The vertical dashed line marks the total SFR of the Milky Way \citep{chomiuk11a}; values to the left of this line, indicated by the arrow, are consistent with the total Galactic SFR, while values to the right of it are inconsistent.
}
}
\end{figure}

{The GCD model has the potential to perform better, since for it $f_{\rm GCD} < 1$, i.e., the star formation rate is potentially lower due to the final dispersal phase in this model. We can address this possibility both analytically and numerically. Analytically, note that \autoref{eq:epsstar_gcd} and \autoref{eq:f_gcd} together imply that
\begin{equation}
f_{\rm GCD} \geq (1 + \eta)\epsilon_*,
\end{equation}
so that values of $f_{\rm GCD} \ll 1$ also imply values of $\epsilon_* \ll 1$, in which case it is difficult to see how bound clusters could form. Indeed, using \autoref{eq:sfr_gcd}, we have
\begin{equation}
\SFR \geq \xi \epsilon_* \left(1 + \tc\right) \left(\frac{M_{\rm tot}}{\tffz}\right) \gtrsim 16.5 \epsilon_*\,M_\odot \mbox{ yr}^{-1},
\end{equation}
where in the numerical evaluation we have set $\xi \geq 1$, since, as noted above, when one uses the spherical equivalent density (as has been done for the ATLASGAL sample), this inequality holds. Thus the observed SFR of the Milky way is only consistent with a GCD model in which $\epsilon_* \lesssim 0.1$, in which case we expect that almost none of the ATLASGAL clumps could go on form a bound cluster. This seems problematic, since if ATLASGAL clumps cannot go on to form bound clusters, it is unclear what structures can.
}

{We can also use our MCMC analysis address the value of $f_{\rm GCD}$, and whether it allows one to simultaneously match the ATLASGAL and YSO count data. To do so, we proceed as follows. First, since we have seen from \autoref{fig:combined_gcd} that ATLASGAL YSO counts essentially constrain only $\eff$, while age distributions constrain other variables but not $\eff$, we select from our MCMC chains for our fit to the stellar age distribution all samples for which $\eff$ lies within the 16th to 84th percentile range allowed by our analysis of YSO counts.\footnote{{We use the fits to NGC 6530 for this purpose, but the results for the ONC are qualitatively the same. Similarly, using values of $\eff$ constrained to lie in the 5th to 95th or the 1st to 99th percentile range also does not change the qualitative result.}} This gives us a set of parameter values that are consistent with both sets of observations. Second, for each sample we compute the quantity $f_{\rm GCD} [(1+\tauc)/\tauc] \eff$ (c.f.~\autoref{eq:sfr_gcd}), and the corresponding predicted value of the SFR for that set of parameters. The result is a set of predicted SFRs for the ATLASGAL clumps, considering only those parameter values that are also consistent with the data on YSO clumps and age distributions.}

{
We plot the distribution of predicted SFRs in \autoref{fig:sfr_pdf}. For comparison, we also plot the corresponding distributions for the GC model (which uses an identical procedure except that $f_{\rm GCD} = 1$ for all samples) and for the CBD model (for which we derive the SFR from \autoref{eq:SFR_CB}). As expected based on the arguments above and on \autoref{eq:SFR_CB} and \autoref{eq:SFR_ATLASGAL}, the CBD model predicts that ATLASGAL clumps form stars at a few tenths of a Solar mass per year, consistent with all of the bound star formation in the Galaxy occurring in them, and perhaps a small amount of unbound as well. The GC model overproduces the SFR of the Galaxy by a factor of $\sim 10$. The figure also shows that the GCD model does not do any better than the GC model at matching the observed Galactic SFR; the extra dispersal at the end, once we constrain the parameters that describe it by the observed age distributions and YSO counts, does not allow a significantly lower total SFR for GCD than for GC. There is a small tail of parameter space that allows the ATLASGAL sample to have a SFR comparable to that of the entire Galaxy, but even this solution is problematic, since these models are viable only to the extent that one is willing to assume that star formation in the Galaxy occurs exclusively in clumps as dense or denser than the ONC, i.e., the lower-density regions like Perseus, Taurus, Ophiuchus, etc., make zero contribution to the Galactic SFR.}

The fundamental problem for {the GC and GCD} model{s} is completely analogous to the one noted by \citet{zuckerman74a} for CO-detected molecular clouds, and by \citet{krumholz07e} for HCN-detected ones: the model assumes that order unity of the mass in the ATLASGAL clumps will be converted to stars on a timescale comparable to the free-fall time, which yields a star formation rate much higher than the one we actually observe in the Milky Way. However, there as an important extra feature here, which is not present in the earlier works. One can avoid the problem of over-producing stars from the CO and HCN data by assuming a very high mass loading factor, 
{
either at all times (in GC) or at late times (in GCD). However, we can now see that this solution is in strong tension with the combined YSO counts and stellar age data. The YSO counts require that the star formation rate per free-fall time stay nearly constant, so the only way for star formation to accelerate, as required by the observed age distributions, is for the total density in the star-forming gas to rise. For the acceleration to be enough to match the observations, this density increase must occur substantially faster than the gas is depleted by star formation or feedback -- in terms of the parameters of our models, we require $\tauc \ll 1$. However, if the density is increasing much faster than gas is removed by feedback, this in turn implies a high total star formation efficiency. There is no way to simultaneously satisfy the constraints of low SFR per free-fall time in individual clumps and accelerating star formation without also overproducing the total SFR of the Galaxy.
}

\section{Summary and conclusion}
\label{sec:conclusion}

In this paper we investigate a number of candidate scenarios for the formation of bound star clusters, focusing on questions of how the mass is assembled, how it evolves, and how efficiently it forms stars. We do so taking advantage of two significant observational advances over the past few years. The first is the availability of spectroscopically-estimated ages for a reasonably complete sample of stars that can be assigned with high confidence to young clusters using \textit{Gaia} kinematics \citep{kounkel18a, prisinzano19a}. These data now show in multiple clusters that star formation in clusters is an accelerating but extended process, i.e., the star formation rate increases over time, but the total duration of star formation is several free-fall times, so that $\sim 30-50\%$ of the stars in any given cluster are more than three free-fall times old, and $\sim 5-10\%$ are as old as ten free-fall times. Explaining this accelerating but extended star formation history requires a model in which either the total mass of gas available for star formation increases with time, the efficiency of star formation at fixed gas mass and density increases with time, or the mean density increases with time, leading to a increase in the star formation rate -- these scenarios roughly correspond to the models of conveyor belt star formation, increasing efficiency of star formation, and global hierarchical collapse that have previously appeared in the literature.

The second data set of which we make use is a large sample of star-forming gas clumps from the ATLASGAL survey \citep{schuller09a, csengeri14a, heyer16a}.  that are well-matched to young star clusters in terms of mass and density, but which are still very gas rich and thus likely represent a slightly earlier evolutionary state. Such gas clumps show a very tight correlation between the mass of gas, its mean density, and the number of young stellar objects (YSOs) embedded within it, which together constrain the rate at which the gas produces YSOs. We carry out a Bayesian forward-modelling treatment of the observational uncertainties and possible biases in these data set, including the effects of selecting only gas-dominated systems and of changes in the gas properties on timescales shorter than the YSO lifetime, and we find that these factors do not significantly alter the overall constraint on how efficiently gas produces YSOs. The tight correlation of gas properties with YSO counts rules out the possibility that the star formation efficiency per free-fall time is time-dependent, ruling out models where the observed acceleration of star formation is due to a time-dependent increase in star formation efficiency per unit mass per unit free-fall time.

We finally consider the global star formation budget of the Milky Way, and show that the scenarios of global hierarchical collapse and conveyor belt star formation predict that the observed ATLASGAL clump population will yield very different total rates of star formation in the Galaxy. The collapse scenario is only able to recover the observed acceleration of star formation if clumps collapse globally on a timescale shorter than that on which they initially form stars locally, since otherwise depletion of the gas by star formation yields a star formation history that decelerates rather than accelerating. However, the requirement for global collapse to occur before a significant fraction of the mass can form stars in turn requires that the ATLASGAL clumps produce stars at a rate that exceeds the entire star formation rate of the Milky Way, let alone the substantially lower rate at which bound star clusters form.

By contrast, the conveyor belt model, first proposed by \citet{longmore14a}, encounters no such difficulties, because it attributes the acceleration of star formation to the fact that gas clumps form stars and accrete simultaneously, so that the gas mass available for star formation tends to increase with time until the gas is dispersed by feedback. We further find that accretion at a rate that varies with time as $\dMa\propto t^3$, as generically predicted for the gravitational collapse of mass reservoirs with fixed bounding pressure \citep{goldbaum11a}, produces a distribution of stellar ages consistent with that observed in young clusters. We therefore conclude that the best available explanation for all of the available observational constraints is that bound star clusters form in a conveyor belt mode, where gas accretes at an increasing rate, but the central cluster-forming region is not in a state of global collapse, and has a star formation efficiency per unit mass that is both low and roughly constant in time.

\section*{Acknowledgements}

{We thank E.~Vazquez-Semadeni, J.~Ballesteros-Paredes, A.~Palau, G.~C.~Gomez, and M.~Zamora-Aviles for comments on the manuscript, and we thank the anonymous referee for a helpful report.} MRK acknowledges funding from the Australian Research Council through the Future Fellowship (FT180100375) and Discovery Projects (DP190101258) funding schemes. CFM acknowledges support by NASA through NASA ATP grant NNX13AB84G. This research made use of Astropy,\footnote{http://www.astropy.org} a community-developed core Python package for Astronomy \citep{astropy-collaboration13a, astropy-collaboration18a}.

\bibliographystyle{mnras}
\bibliography{refs}

\clearpage

\appendix

\section{Supplementary Material (online-only)}
\label{app:supp_figures}

\begin{figure*}
\includegraphics[width=\textwidth]{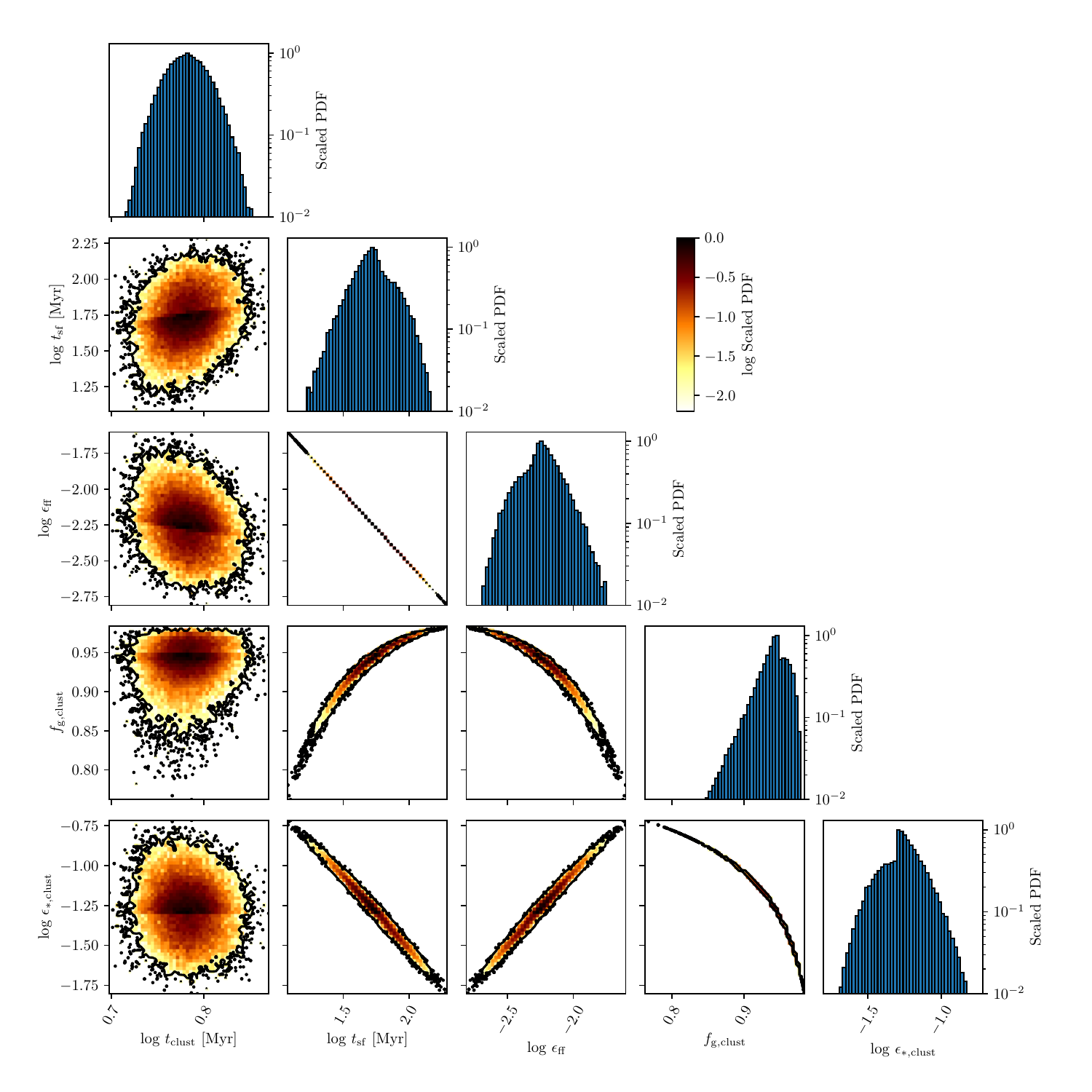}
\caption{
\label{fig:ONC_ST}
Corner plot showing the distributions of both directly fit parameters and derived parameters from our MCMC application of model ST to the stellar age distribution in the ONC. The central panels show heat maps of the probability density in two dimensional cuts through the indicated axes, each scaled to have a maximum of unity; the outer contour marks a scale probability density of $0.01$, and individual points outside this contour correspond to individual MCMC samples. Panels containing histograms show the marginal posterior probability distributions for each parameter. The parameters shown are the same as those listed in \autoref{tab:fit_params}, and are computed in the same way, i.e., only $\tobs$ and $\tsf$ are fit as part of the MCMC, while all other quantities are derived from them; this is why, for example, $\eff$ and $\tsf$ are perfectly correlated.
}
\end{figure*}

\begin{figure*}
\includegraphics[width=\textwidth]{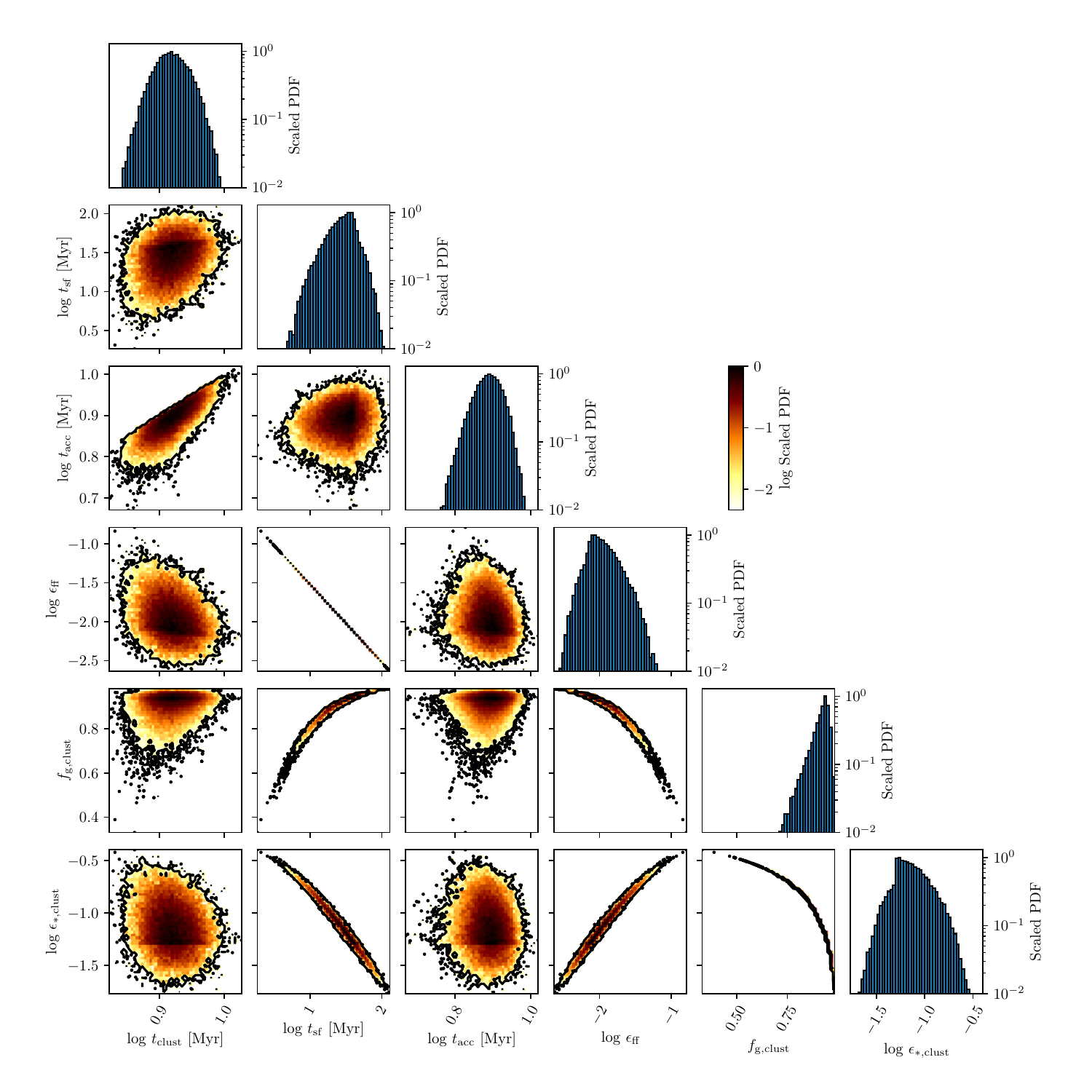}
\caption{
\label{fig:ONC_CBp0}
Same as \autoref{fig:ONC_ST}, but for model CB with $p=0$.
}
\end{figure*}

\begin{figure*}
\includegraphics[width=\textwidth]{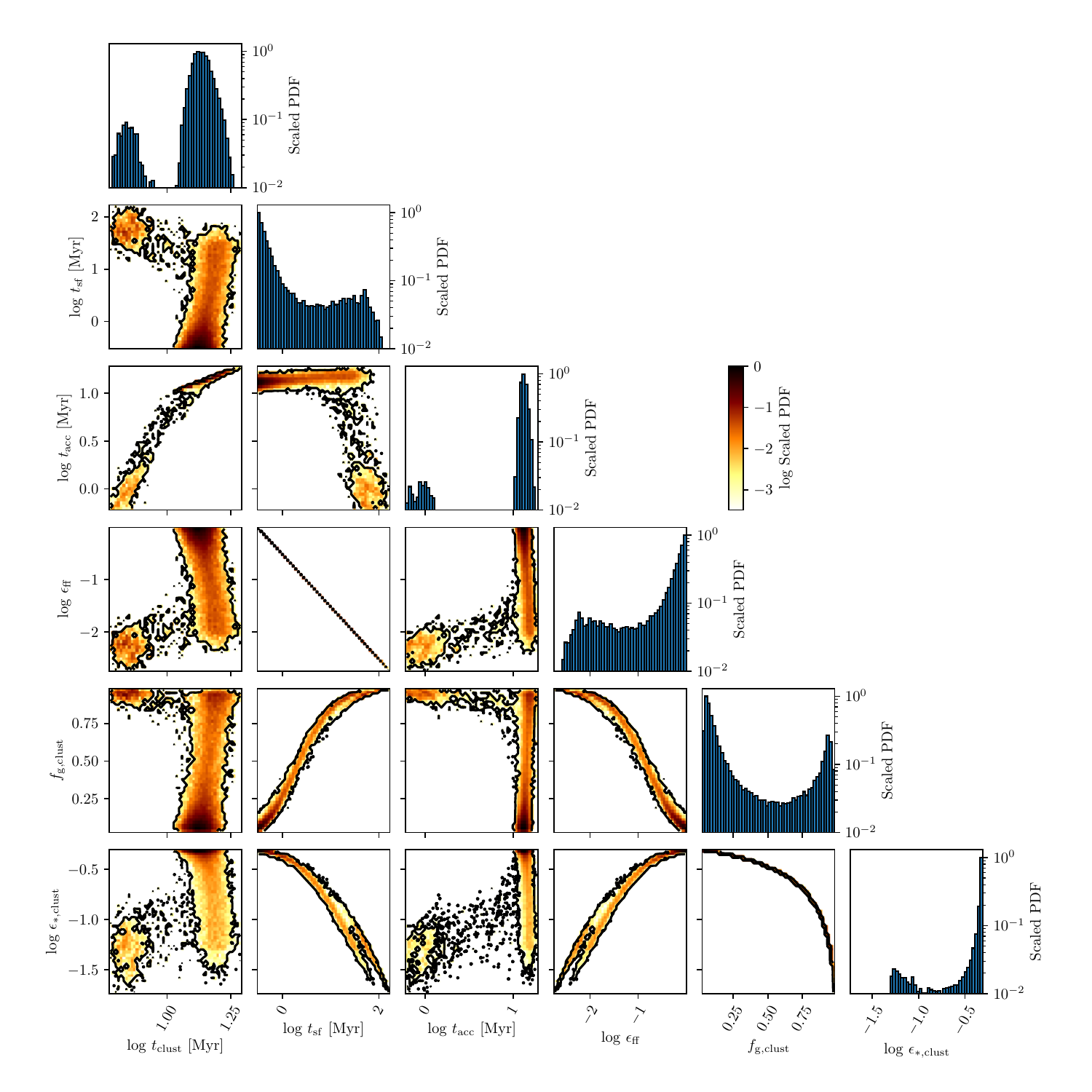}
\caption{
\label{fig:ONC_CB}
Same as \autoref{fig:ONC_ST}, but for model CB with $p=3$.
}
\end{figure*}

\begin{figure*}
\includegraphics[width=\textwidth]{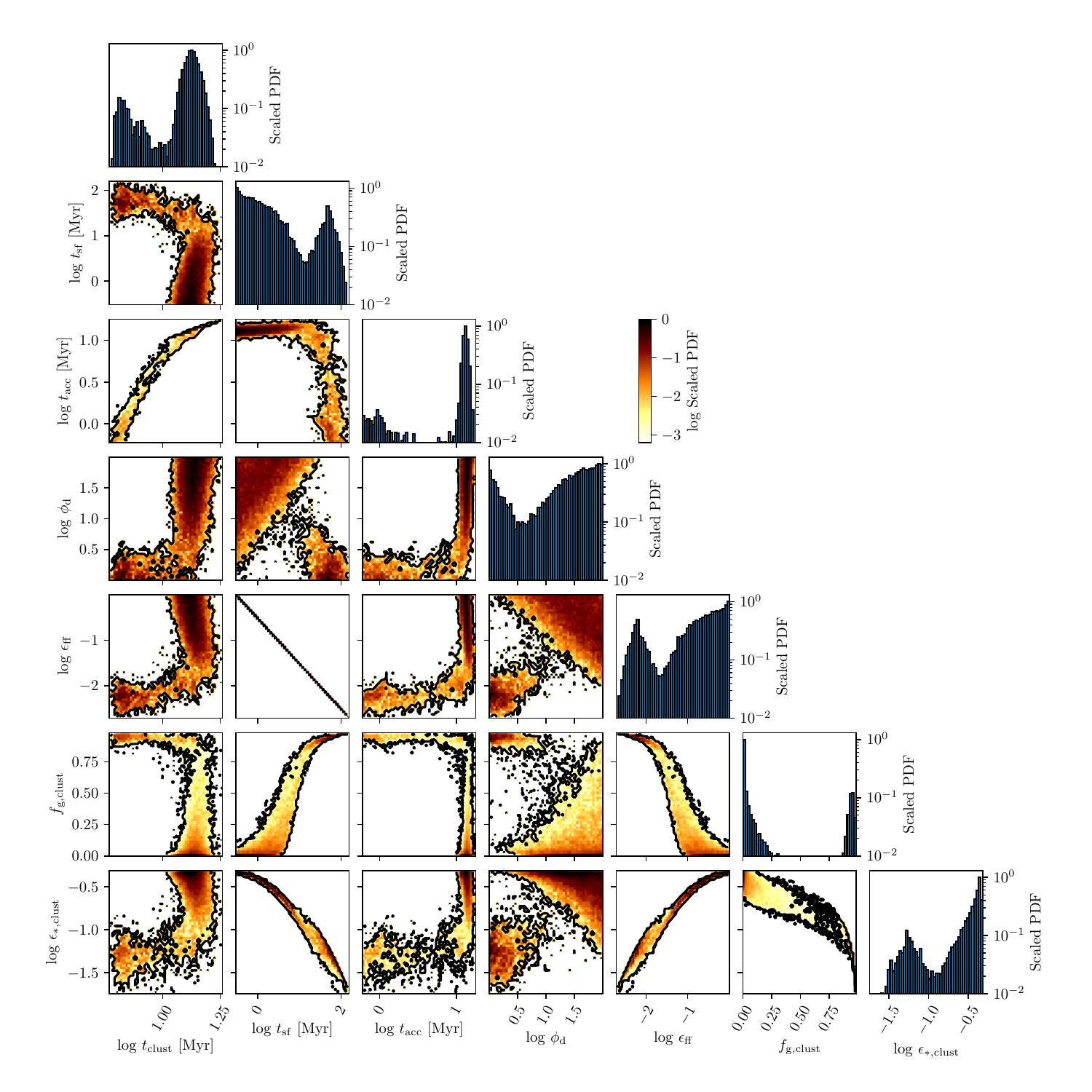}
\caption{
\label{fig:ONC_CBD}
Same as \autoref{fig:ONC_ST}, but for model CBD with $p=3$.
}
\end{figure*}

\begin{figure*}
\includegraphics[width=\textwidth]{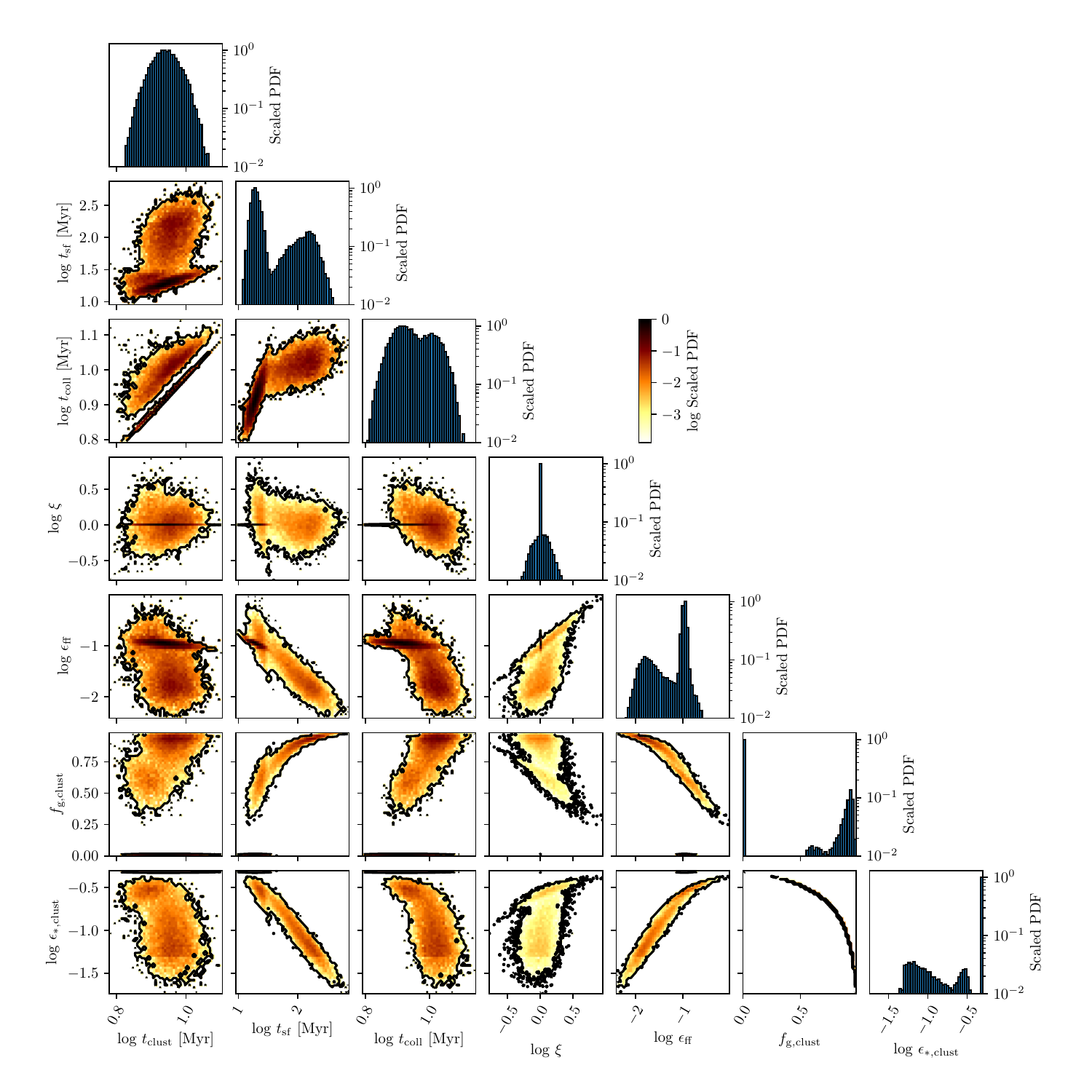}
\caption{
\label{fig:ONC_GC}
Same as \autoref{fig:ONC_ST}, but for model GC. Note that $\xi$ is not directly fit, but is derived from the fit parameters. However, we provide it for convenience.
}
\end{figure*}

\begin{figure*}
\includegraphics[width=\textwidth]{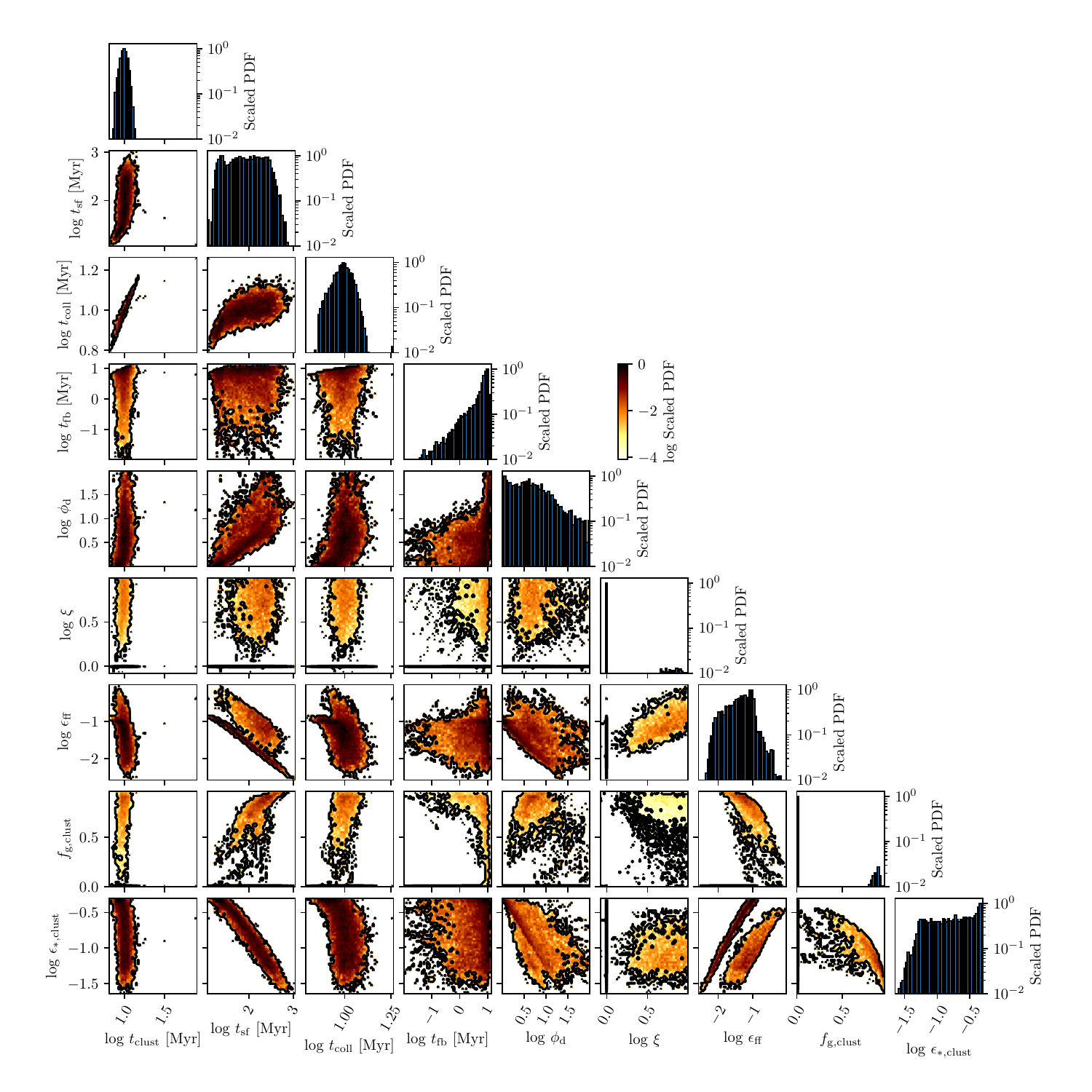}
\caption{
\label{fig:ONC_GCD}
{Same as \autoref{fig:ONC_ST}, but for model GCD. Note that $\xi$ is not directly fit, but is derived from the fit parameters. However, we provide it for convenience.}
}
\end{figure*}

\begin{figure*}
\includegraphics[width=\textwidth]{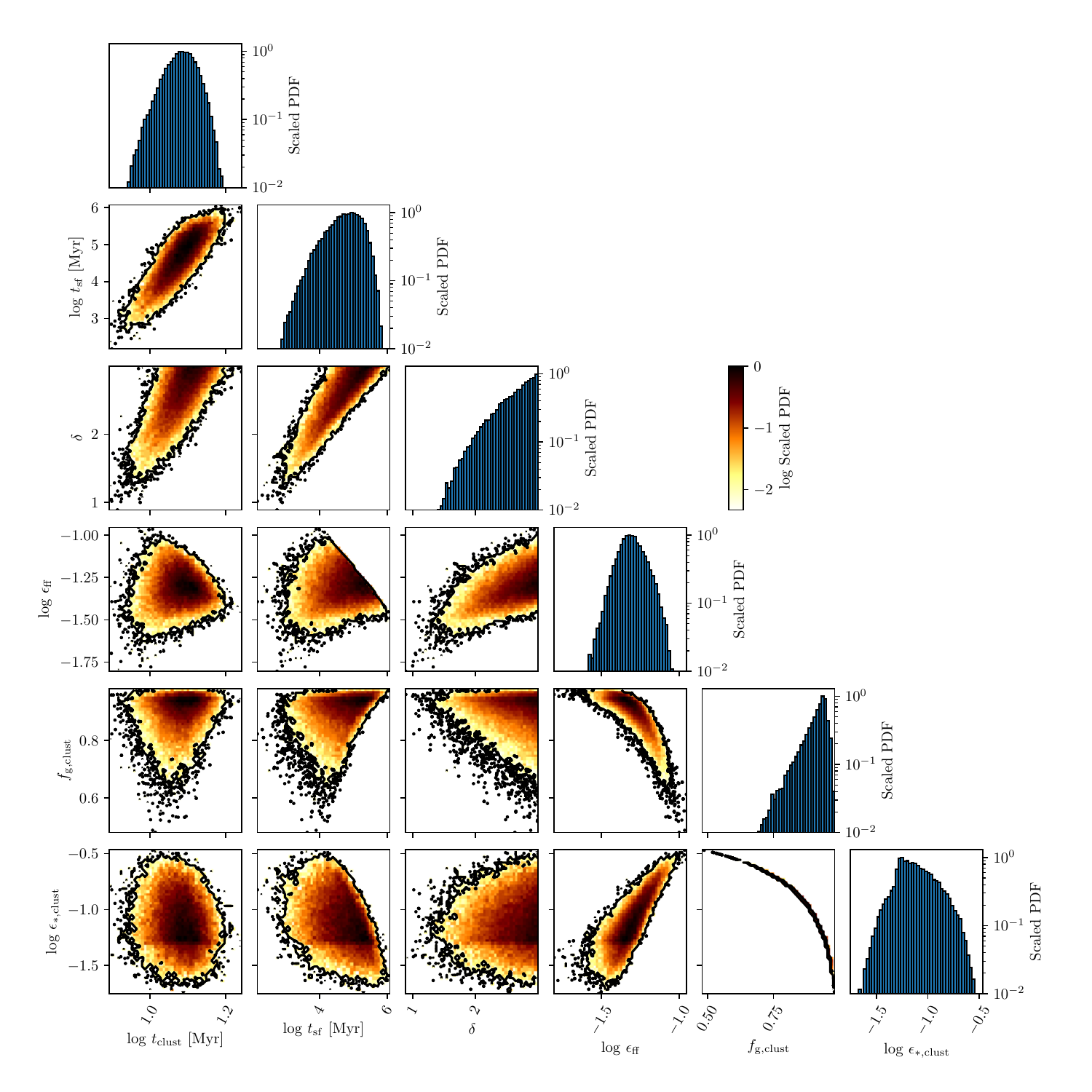}
\caption{
\label{fig:ONC_IE}
Same as \autoref{fig:ONC_ST}, but for model IE.
}
\end{figure*}

\begin{figure*}
\includegraphics[width=\textwidth]{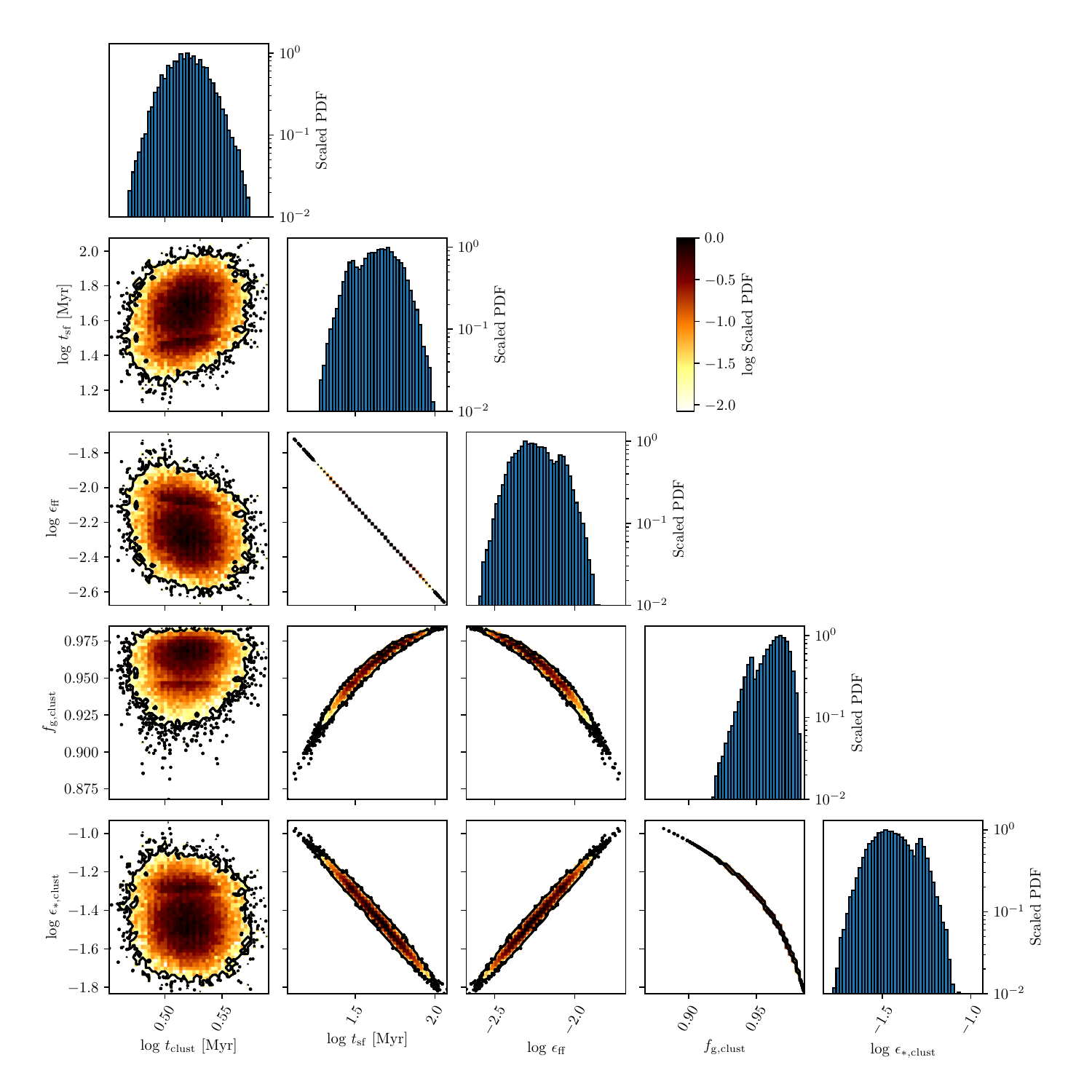}
\caption{
\label{fig:NGC6530_ST}
Same as \autoref{fig:ONC_ST}, but for NGC 6530 rather than the ONC.
}
\end{figure*}

\begin{figure*}
\includegraphics[width=\textwidth]{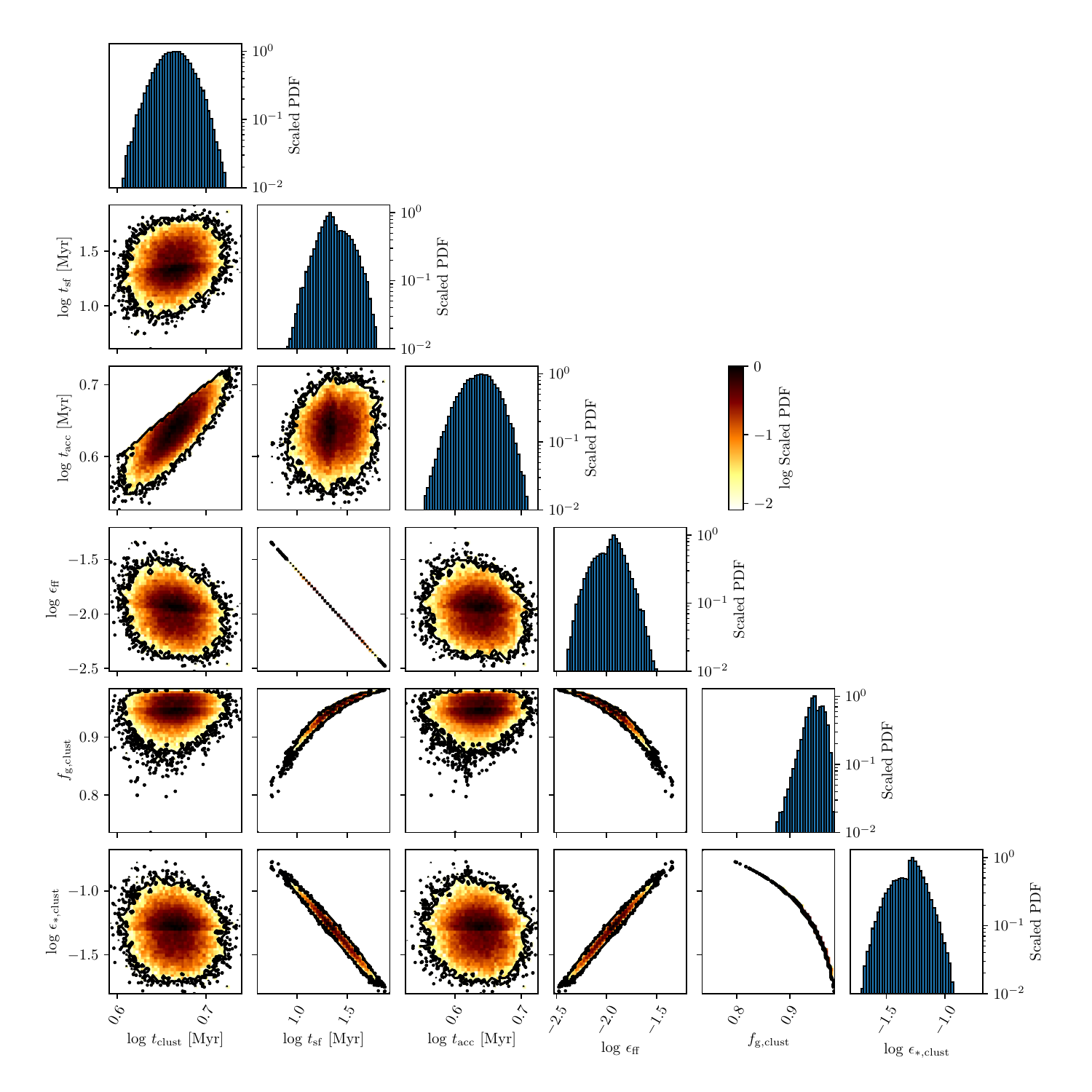}
\caption{
\label{fig:NGC6530_CBp0}
Same as \autoref{fig:NGC6530_ST}, but for model CB with $p=0$.
}
\end{figure*}

\begin{figure*}
\includegraphics[width=\textwidth]{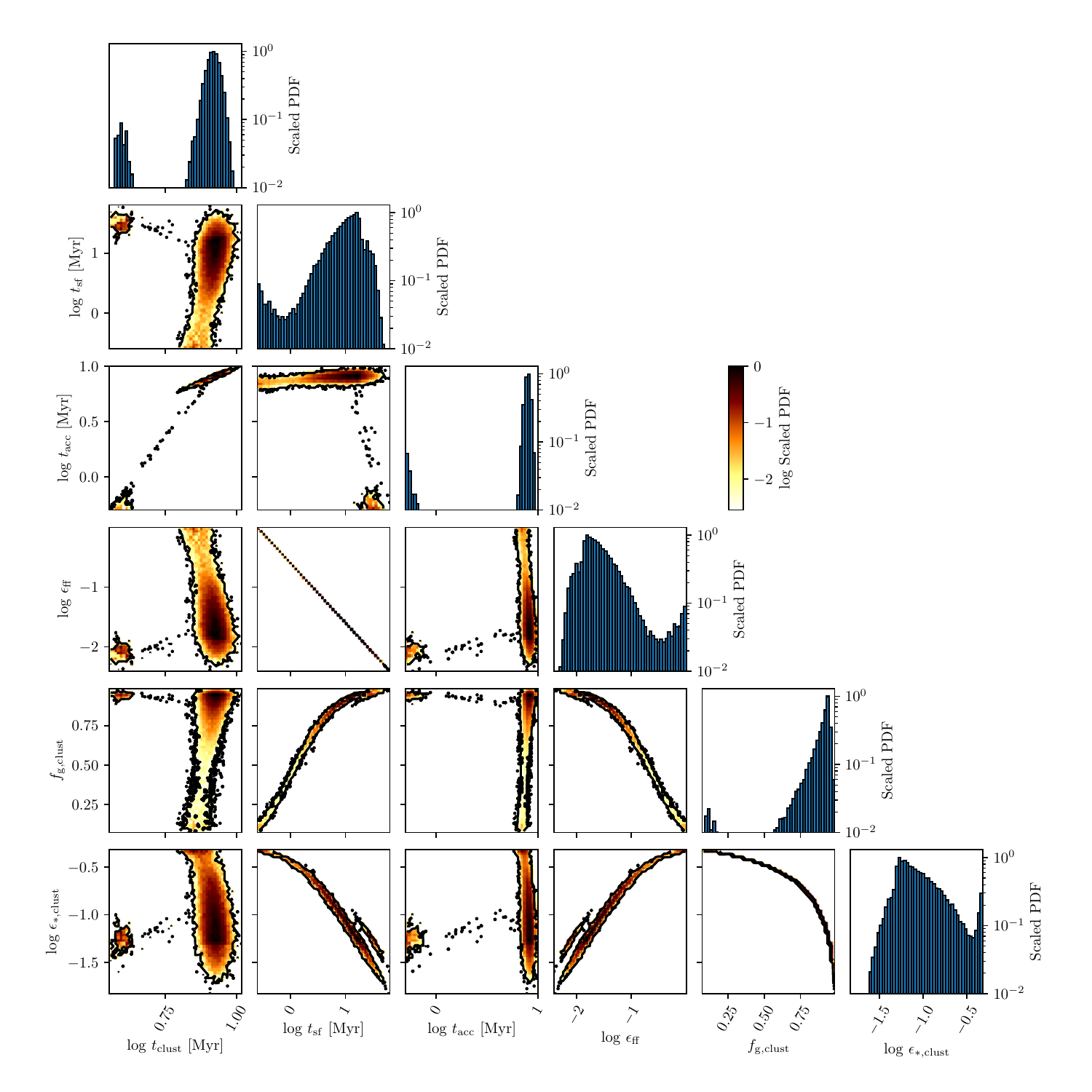}
\caption{
\label{fig:NGC6530_CB}
Same as \autoref{fig:NGC6530_ST}, but for model CB with $p=3$.
}
\end{figure*}

\begin{figure*}
\includegraphics[width=\textwidth]{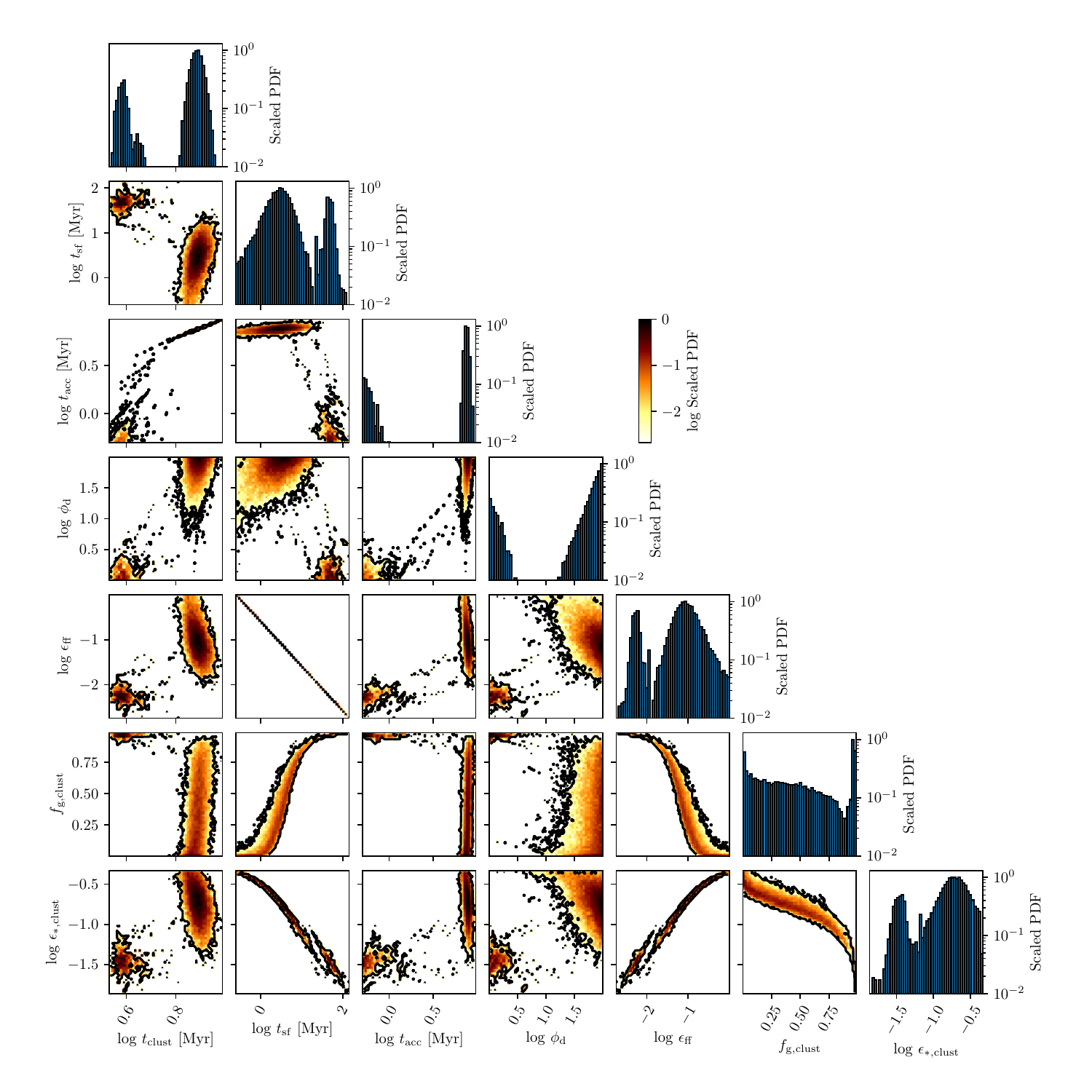}
\caption{
\label{fig:NGC6530_CBD}
Same as \autoref{fig:NGC6530_ST}, but for model CBD with $p=3$.
}
\end{figure*}

\begin{figure*}
\includegraphics[width=\textwidth]{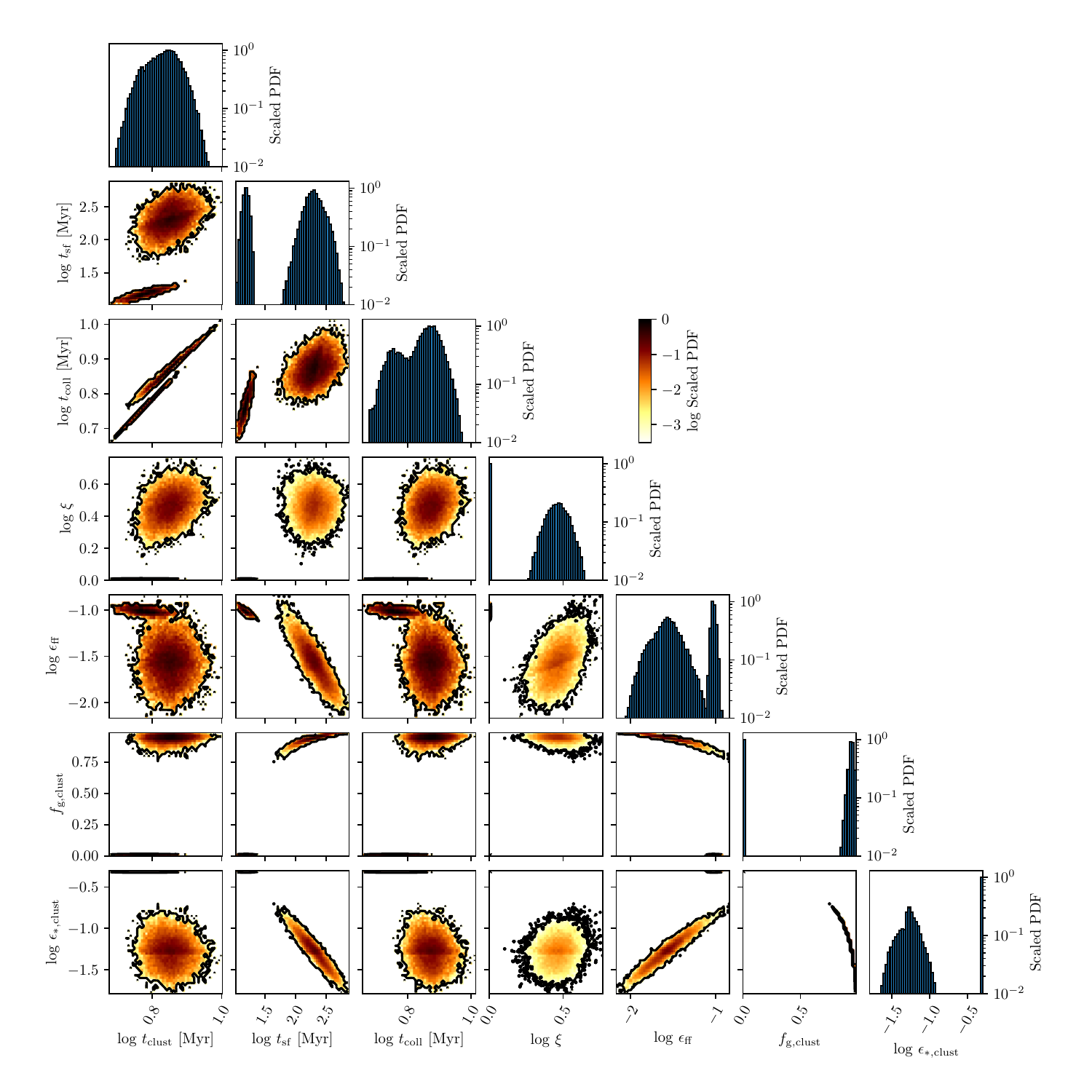}
\caption{
\label{fig:NGC6530_GC}
Same as \autoref{fig:NGC6530_ST}, but for model GC. Note that $\xi$ is not directly fit, but is derived from the fit parameters. However, we provide it for convenience.
}
\end{figure*}

\begin{figure*}
\includegraphics[width=\textwidth]{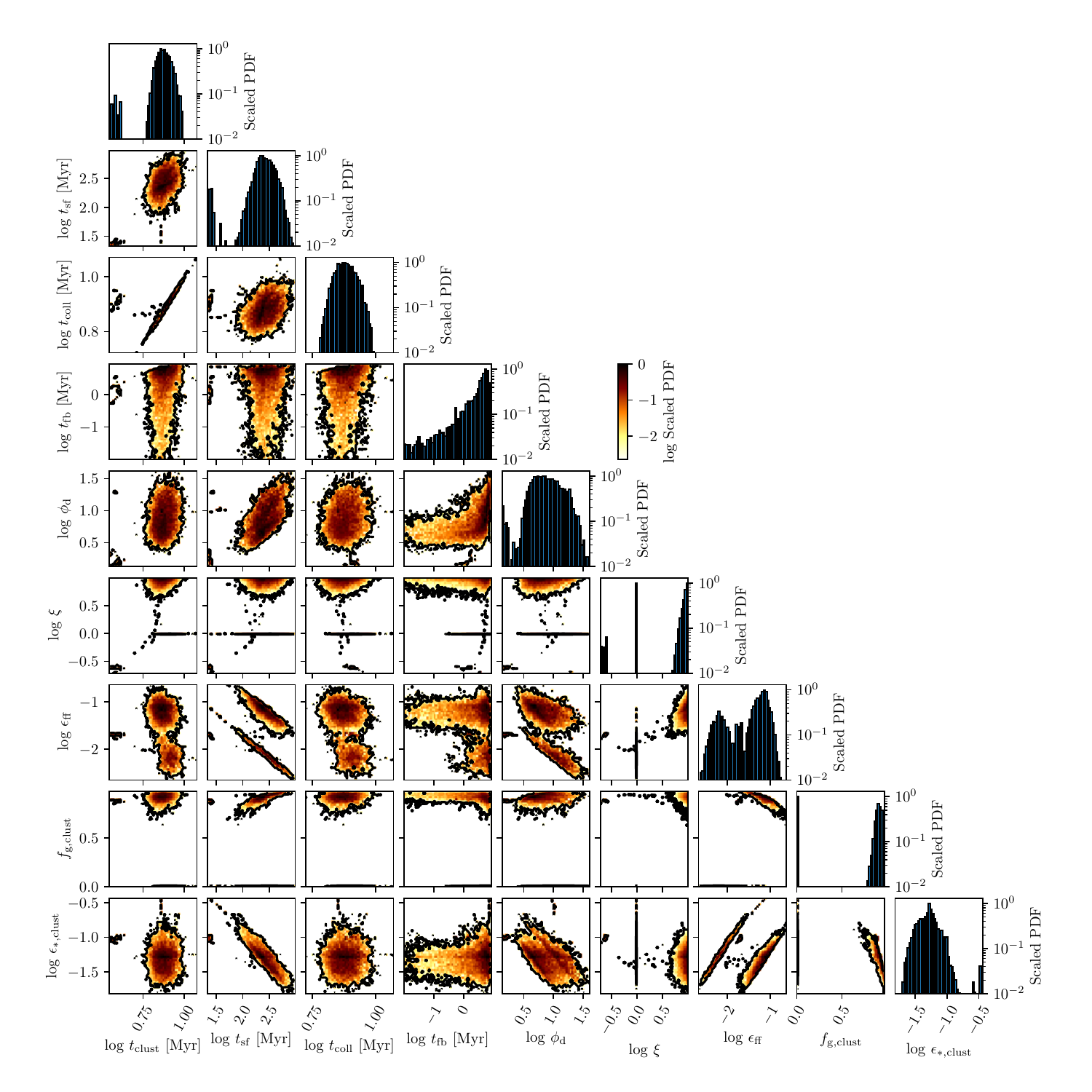}
\caption{
\label{fig:NGC6530_GCD}
{Same as \autoref{fig:NGC6530_ST}, but for model GC. Note that $\xi$ is not directly fit, but is derived from the fit parameters. However, we provide it for convenience.}
}
\end{figure*}

\begin{figure*}
\includegraphics[width=\textwidth]{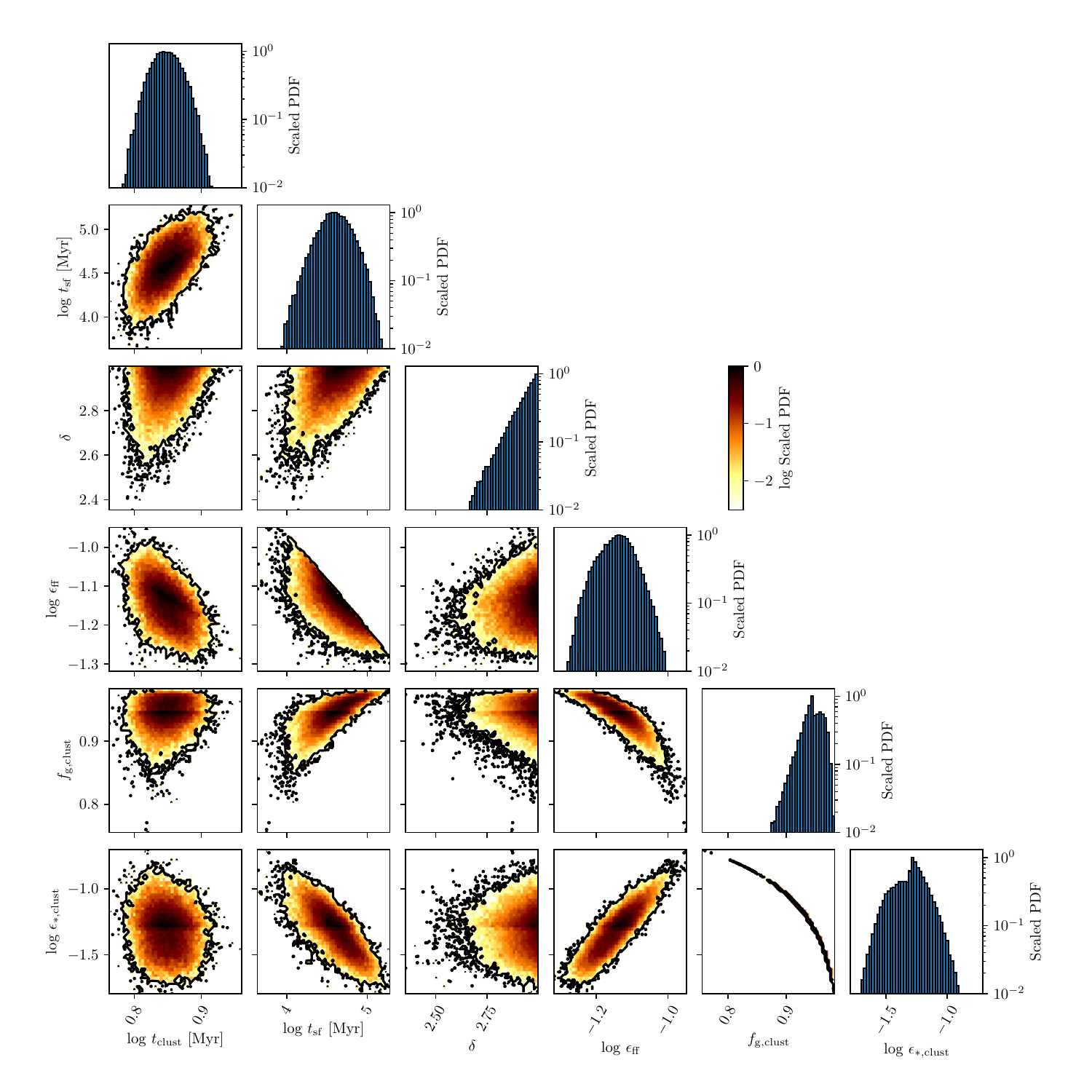}
\caption{
\label{fig:NGC6530_IE}
Same as \autoref{fig:NGC6530_ST}, but for model IE.
}
\end{figure*}

\begin{figure}
\includegraphics[width=\columnwidth]{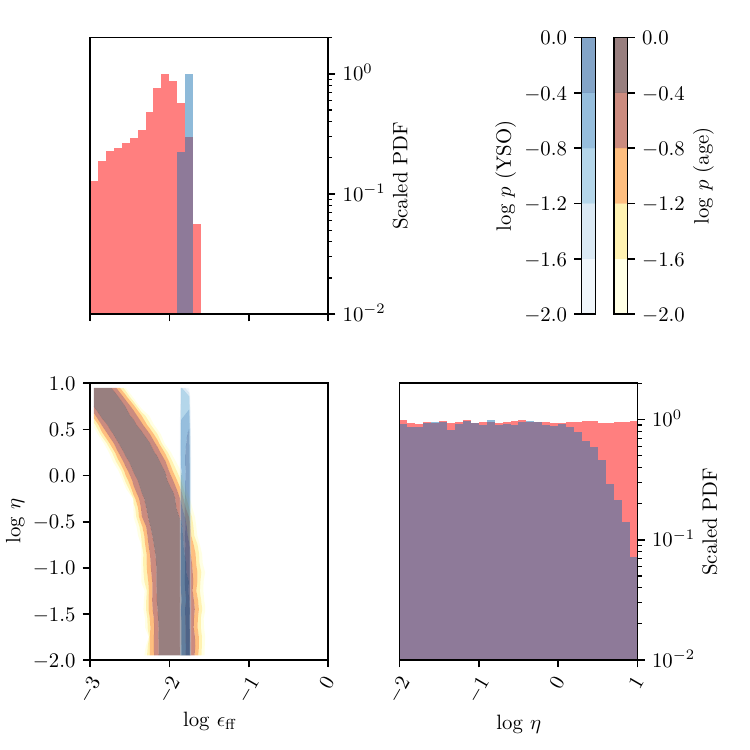}
\caption{
\label{fig:combined_st}
Same as \autoref{fig:combined_cbd} of the main text, but for the ST model.
}
\end{figure}

\begin{figure}
\includegraphics[width=\columnwidth]{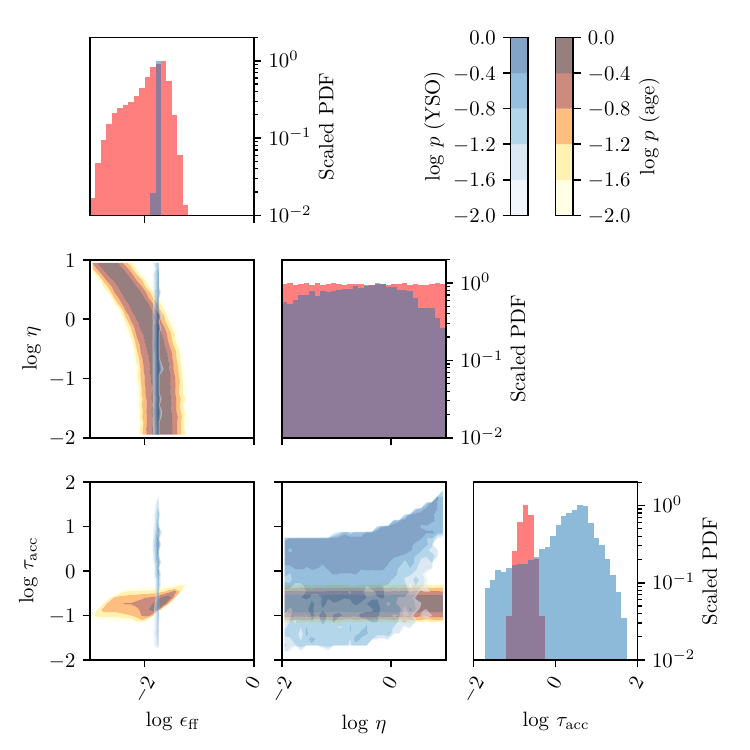}
\caption{
\label{fig:combined_cbp0}
Same as \autoref{fig:combined_cbd} of the main text, but for the CB ($p=0$) model.
}
\end{figure}

\begin{figure}
\includegraphics[width=\columnwidth]{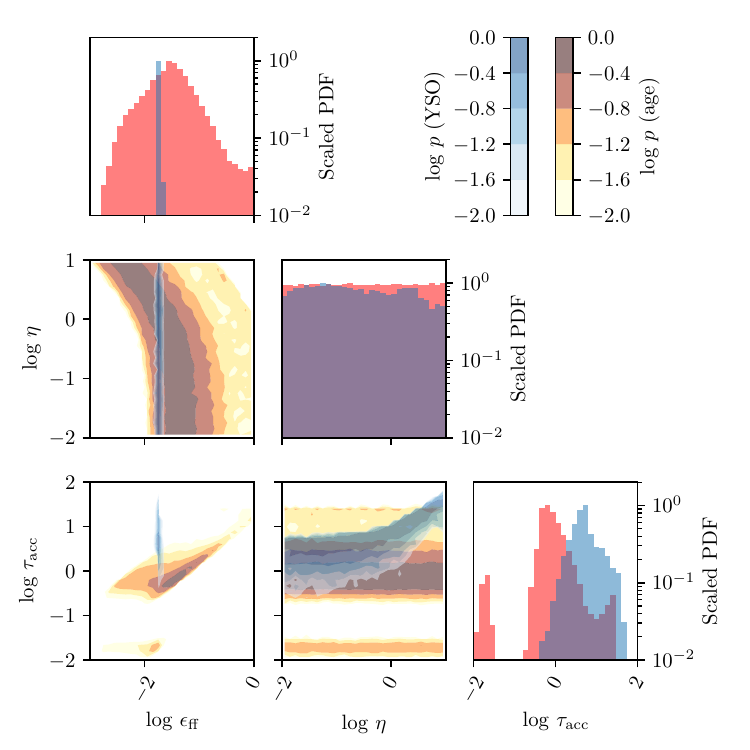}
\caption{
\label{fig:combined_cb}
Same as \autoref{fig:combined_cbd} of the main text, but for the CB model.
}
\end{figure}

In Figures \ref{fig:ONC_ST} - \ref{fig:NGC6530_IE}, we provide full posterior PDFs resulting from our MCMC fits of all models to the stellar age distributions in the ONC and NGC 6530. These plots are all for our fiducial errors $(\sigma,b) = (0.12, -0.05)$ dex, and the PDFs shown are derived using samples from the final 400 iterations of the MCMC. In Figures \ref{fig:combined_st} - \ref{fig:combined_cb} we provide the full combined posterior PDFs derived from stellar age distributions and YSO counts in ATLASGAL clumps for the ST, CB ($p=0$), and CB models.

\begin{figure}
\includegraphics[width=\columnwidth]{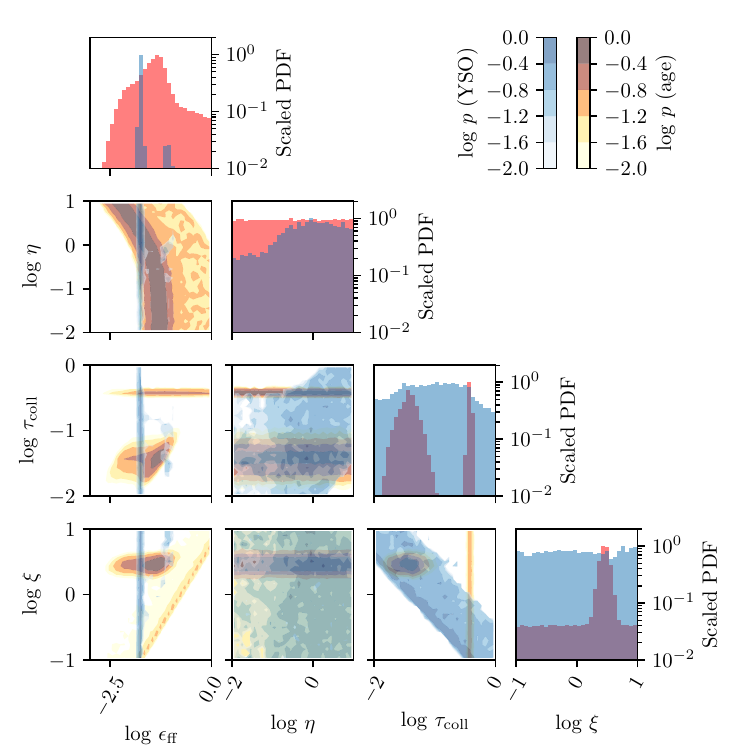}
\caption{
\label{fig:combined_gc}
Same as \autoref{fig:combined_cbd} of the main text, but showing the GC model.
}
\end{figure}

\bsp	
\label{lastpage}
\end{document}